\documentclass[twocolumn]{aastex63}

\usepackage{comment}
 \usepackage{hyperref}
\usepackage{amsmath}
\usepackage{xcolor,xfrac,hyperref,datetime}

 \received{\today:\currenttime}

\shorttitle{Metastable helium}
 \shortauthors{Kulkarni}
 
\graphicspath{{./}{Figures/}}

\begin{document}

\title{Metastable helium in the thermosphere}

\correspondingauthor{S.\ R. Kulkarni}
 \email{srk@astro.caltech.edu}

\author[0000-0001-5390-8563]{S.\ R.\ Kulkarni}
 \affiliation{Owens Valley Radio Observatory 249-17,
 California Institute of Technology, 
 Pasadena, CA 91125, USA}

\begin{abstract}

SPHEREx, a recently launched astronomy mission, detected a bright
1.083\,$\mu$m emission feature in the commissioning data. The PI
group attributed this feature to the He~I 1.0833\,$\mu$m triplet
line.  Here, I review the physics and aeronomy of this well-known
line of atmospheric origin.  SPHEREx is in a dawn-dusk sun-synchronous
polar orbit, circling the earth nearly 15 times a day and observing
close to the terminator plane.  With a height of 650\,km, SPHEREx
is located in the upper thermosphere that is dominated by atomic
oxygen and helium.  The He~I line is a result of resonance scattering
of solar photons by metastable helium atoms.  It appears that SPHEREx
has the capacity to provide a rich dataset (global, daily, and
2-minute cadence) of the column density of metastable helium in the
upper thermosphere.  As an example of this assertion, {\it  with
data from just one orbit},  the winter helium bulge was readily
seen.  Rapid variations in the column density of metastable helium
is seen over the south pole, which is probably due to spatial
structure in the distribution of metastable helium as well as solar
activity. Helium in the thermosphere is of considerable interest
to operators of low-earth orbiting (LEO) satellites, since drag in
the thermosphere is the primary cause of the decay of these satellites.
SPHEREx, along with on-going ground-based studies (passive NIR
spectroscopy, lidar, incoherent scatter radar), is poised to
contribute to this topic.
 \\
\end{abstract}

\section{SPHEREx}
 \label{sec:SPHEREx}
 
On March 12, 2025, SPHEREx\footnote{\url{https://spherex.caltech.edu/}}
was launched into a polar orbit from Vandenberg Space Force Base.
The goal of this mission is to image the entire sky in 102
low-resolution ``spectrophotometric" bands covering the wavelength
range 0.75--5\,$\mu$m.  The planned survey began on May 1, 2025 and
amazingly enough, only two months later, calibrated data, available
to the astronomical community, started to flow to the NASA/IPAC
Infrared Science
Archive\footnote{\url{https://www.ipac.caltech.edu/project/irsa}}
(IRSA). New data, with a delay of two months, are added to this
archive.

\cite{cwa+20} provides a comprehensive technical description of the
mission. Briefly, a 20-cm telescope feeds six HAWAII-2RG detectors
(cf.\ \citealt{H11}).  Each detector, preceded by a linear variable
filter, covers a portion of the full wavelength band of the instrument.
For instance, ``band~1" covers the wavelength range 0.75--1.1\,$\mu$m
whilst band~2 covers 1.10--1.62\,$\mu$m and so on.  Scanning of the
sky results in a given sky position sampling different parts of the
bandpass.  Following in-orbit check-out, the PI team identified a
bright spectral line which was seen in all band~1 images with the
He~I\,1.0833\,$\mu$m triplet.  This line has been known to atmospheric
scientists through ground-based observations since 1957.  In the
atmosphere of the earth it arises from resonance scattering of solar
photons by helium atoms that are in the 1s2s~${\rm ^3S_1}$ level.
Helium atoms in the 1s2s~${\rm ^3S_1}$ level decay extremely slowly
to the ground state (1s$^2$\,${\rm ^1S_0}$), whence the moniker,
``metastable" helium.

SPHEREx obtains images in six bands, at a uniform cadence (integration
time of 113.6\,s).  The thermospheric He~I intensity data will
certainly provide grist for the aeronomer's mill.  Many of my
astronomy colleagues have been complaining that this He~I line is
stymieing their quest to understand the rest of the universe.  Some
have been tempted to apply AI/ML to predict or minimize the scourge.
A modest understanding of the underlying physics and phenomenology
provides a strong foundation for building such models.  Finally,
tyros and neophytes of the interstellar medium (ISM) would benefit
from understanding the phenomena and physics of this important line
but in a very different physical setting.

The remainder of this paper is organized as follows.  In
\S\ref{sec:Observations}, I summarize the band~1 SPHEREx observations.
In \S\ref{sec:MetastableHelium}, I recount the history of metastable
helium. This is followed by a brief summary of the use of the
principal lines of metastable helium (1.0833\,$\mu$m and $\lambda$3889)
in astronomy and aeronomy.  In \S\ref{sec:Thermosphere} the
(astronomical) reader is presented with a summary of the vertical
structure of the atmosphere with emphasis on helium.
\S\ref{sec:MetastableHeliumThermosphere} is focused on metastable
helium in the thermosphere.  The concentration of metastable helium
in the thermosphere is determined by a number of processes (the
physics of which is well known but the aeronomy of which still
constitutes state-of-the-art research).  For this reason, the column
density of metastable helium in the upper thermosphere (and beyond)
is treated as a free parameter (whose value to be determined by
observations).  In \S\ref{sec:ResonanceScattering}, assuming that
the primary source of He~I photons from resonance scattering of
solar photons by metastable helium, I computed the brightness of
the He~I line(s). Good agreement is found between the calculations
and deductions from recent ground-based (lidar) observations. A
summary of possible destruction pathways for metastable helium in
the upper thermosphere concludes this section.  Since the brightness
of the He~I lines is directly proportional to the intensity of
sunlight, it is important that the reader has a clear understanding
of the intensity of sunlight along the orbit of SPHEREx.  With this
in mind, the orbital geometry is summarized in \S\ref{sec:OrbitalGeometry}.
In \S\ref{sec:Variations} I present He~I line intensity for two
full days, centered on the northern solstice (2025 June 20 and 21).
The observed strong asymmetry between He~I emission in the northern
and southern ecliptic poles is attributed to a manifestation of the
``winter helium bulge". Dramatic variations in He~I line intensity,
on minute timescales, in the south pole region. This is likely a
result of spatial variations as well as solar activity.  The
continuous cadence and comprehensive sampling of the thermosphere
close to the day-night terminator plane will result in a unique
data set for aeronomers (\S\ref{sec:WayForward}). I conclude this
section discussing possible science returns for aeronomy by scheduling
observations with ground-based facilities at times of overhead
passage of SPHEREx.

\begin{figure}[htbp]     
 \plotone{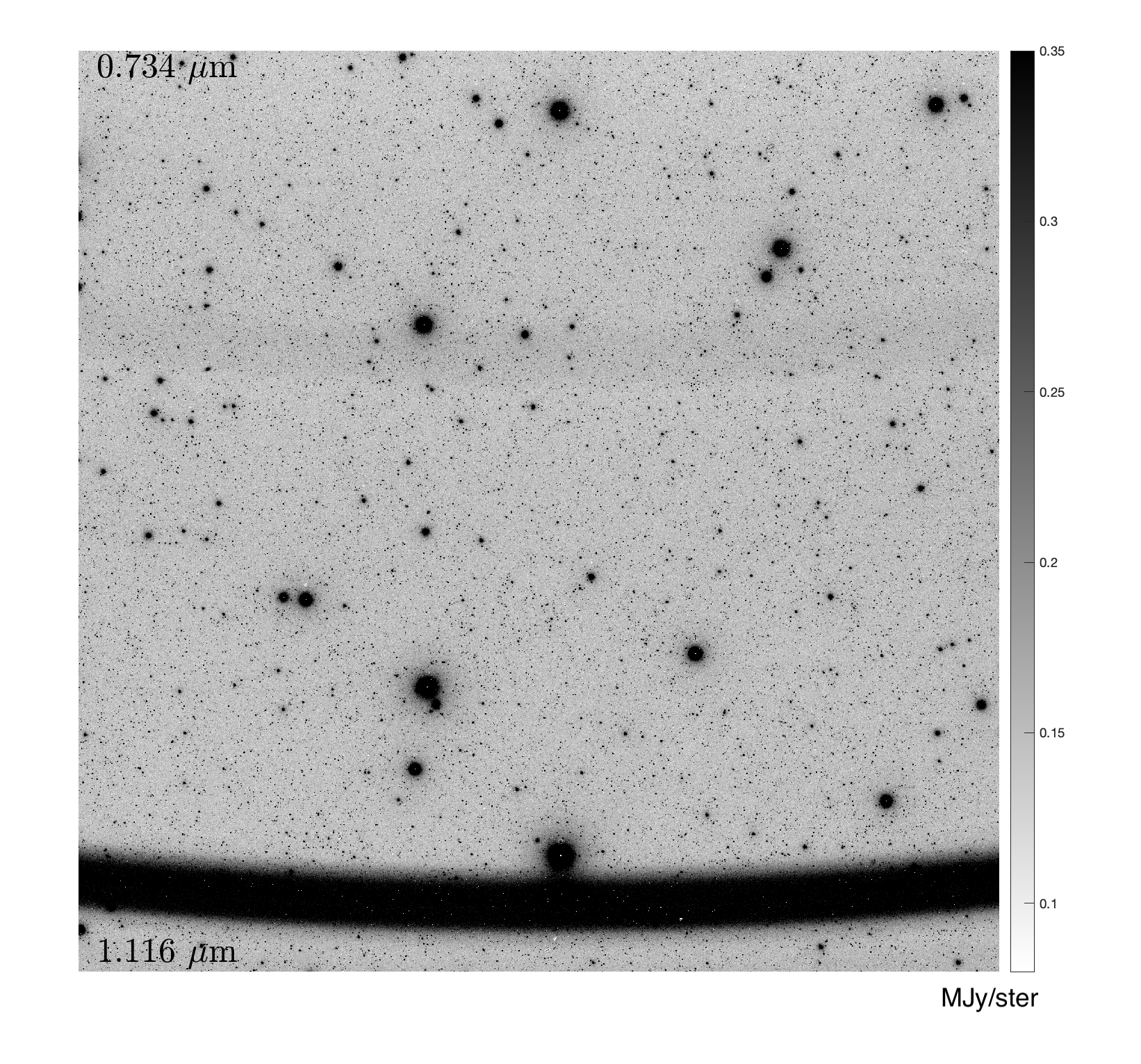}
  \caption{\small Band~1 image of the sky taken at UT 2025-06-20
  00:00:46.9. The image size is $3.5^\circ\times 3.5^\circ$; North
  is up and East to the left. The pixel values are in MJy/ster and
  the thermometer (right side) provides the gray-scale mapping to
  pixel values.  The linear variable filter is designed to provide
  a gradient in the pass-band wavelength in the vertical direction.
  For this band, as noted in the figure, the passband wavelength
  at the top is 0.734\,$\mu$m, increasing to $1.116\,\mu$m towards
  the bottom. The unit for the  numbers on the (right) vertical
  thermometer  is MJy/ster. The dark slightly curved horizontal
  band at the bottom of the image is the atmospheric He~I 1.0833\,$\mu$m
  triplet. The ghostly curved band at 0.85\,$\mu$m is an artifact
  (see \S\ref{sec:DataAnalysis}). }
 \label{fig:band1_image}
\end{figure}

Unless otherwise mentioned, all atomic data are from the Atomic
Spectrum
Database\footnote{\url{https://www.nist.gov/pml/atomic-spectra-database}}
of NIST.  All wavelengths are in vacuum.

\section{The Observations}
	\label{sec:Observations}

\begin{figure}[htbp]   
 \plotone{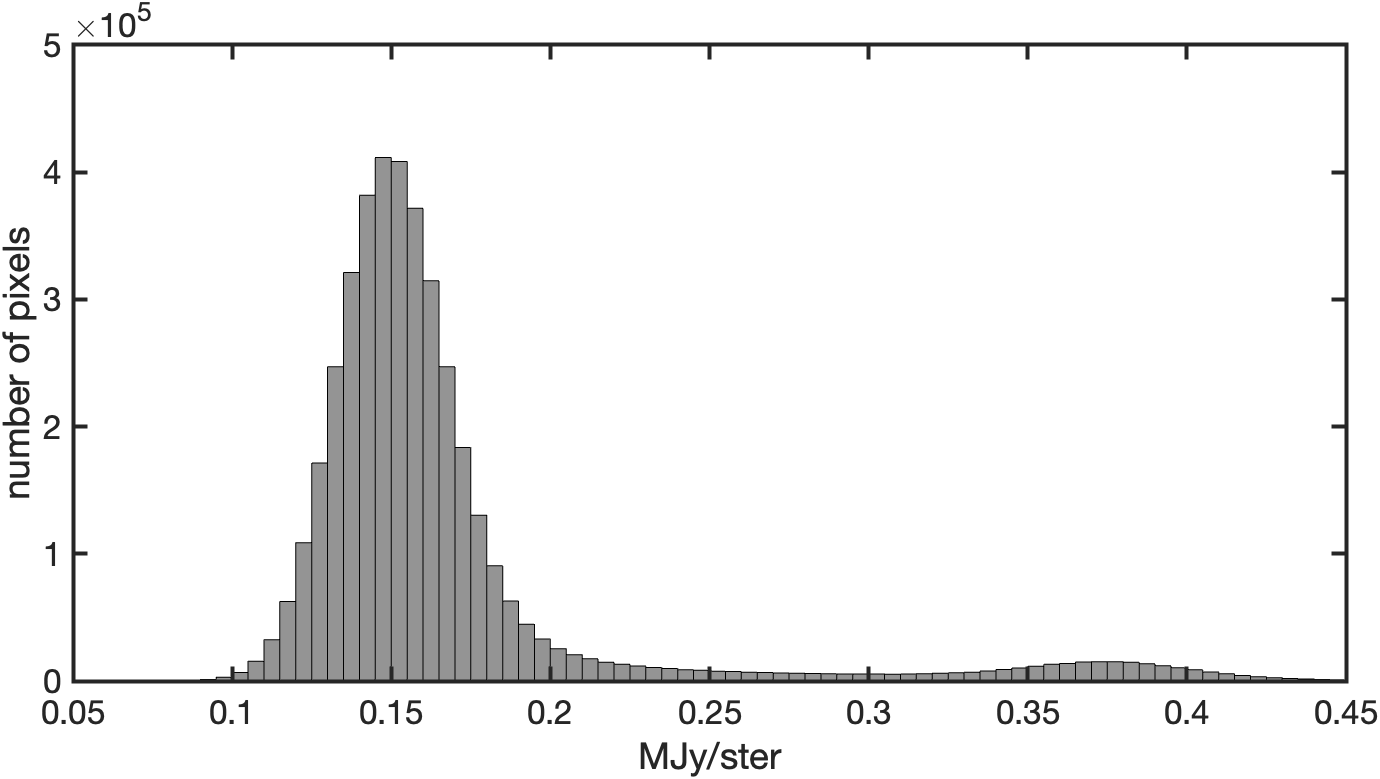}
  \caption{\small Histogram of the pixel values for the image shown
  in Figure~\ref{fig:band1_image}.  The zodiacal background is
  measured by the mode. The secondary hump at 0.38\,MJy\,ster$^{-1}$
  is due to terrestrial He~I\,1.0833\,$\mu$m emission.  }
 \label{fig:band1_histogram}
\end{figure}

The fundamental reference to SPHEREx data is the ``Explanatory
Supplement"\footnote{
\url{https://irsa.ipac.caltech.edu/data/SPHEREx/docs/SPHEREx_Expsupp_QR_v1.1.pdf}}
(hereafter, the {\it Supplement}) provided by IRSA.  Band~1 data
were downloaded from IRSA.  An example band~1 image  is shown in
Figure~\ref{fig:band1_image}.  Note the vertical gradient in
$\lambda$, the passband wavelength.  The spectral resolution,
$\mathcal{R}=\lambda/\delta\lambda$, is determined by the vertical
gradient; here, $\delta\lambda$ is the full width at half-maximum
of the linear filter response function for a narrow line. For band~1,
$\mathcal{R}=39$.  The bright curved feature at the bottom of the
image is centered on 1.083\,$\mu$m and identified with He~I\,1.0833\,$\mu$m
triplet.   The histogram of the pixel values is presented in
Figure~\ref{fig:band1_histogram}.

\begin{figure}[htbp]   
 \centering
  \includegraphics[width=3in]{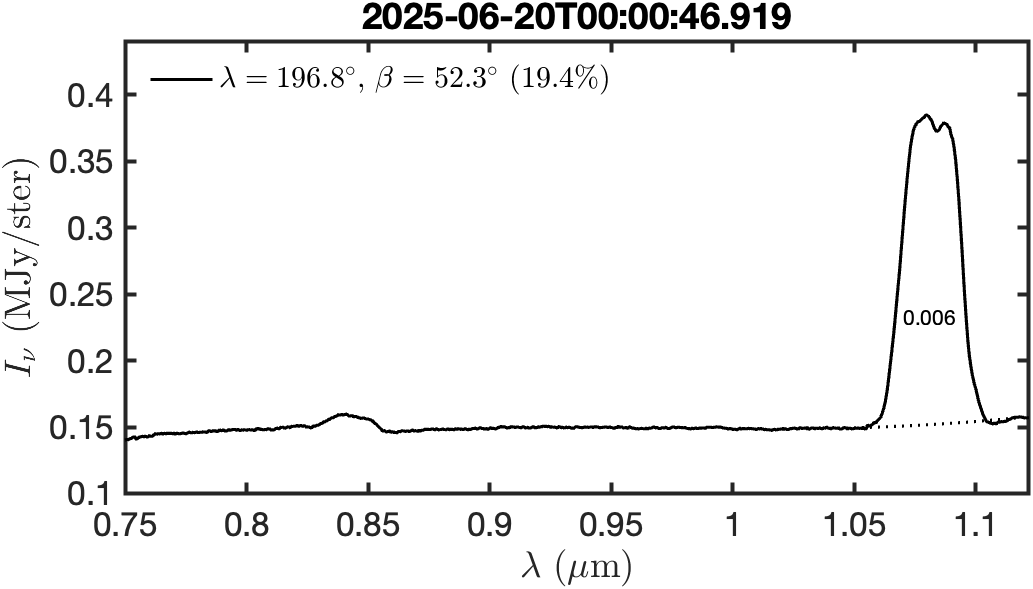}
   \caption{\small Sky-spectrum (solid black line) of band~1 image
   shown in Figure~\ref{fig:band1_image}. The He~I triplet is the
   bright feature at the right end of the spectrum. The dotted black
   line is the baseline constructed from the wavelength range
   $[1.02,1.05]\,\mu$m and $[1.115,1.16]\,\mu$m.  The full-width
   at half-maximum of the He~I line is found to be 0.026\,$\mu$m,
   which is in accord with the pre-launch specifications for the
   instrument \citep{cwa+20}.  $F$,  the integral of the intensity
   over the line  (unit: MJy\,ster$^{-1}$\,$\mu$m), is noted above
   the black dotted line. The geo-centric ecliptic longitude
   ($\lambda$) and latitude ($\beta$) of the bore-sight  are noted
   in the legend.  Also included in the legend is the percentage
   pixels rejected in the image frame (see \S\ref{sec:DataAnalysis}).
   The sun-earth(center)-satellite angle, at the start of  the
   observation (shown in title as UT time) is 76$^\circ$ and the
   sun-earth(center)-bore-sight angle is 97$^\circ$.  The integration
   time is 113.6\,s. The dimple in the middle of the He~I line is
   an artifact anticipated from pre-flight model and the small bump
   at 0.85\,$\mu$m is due to the ghost feature seen in
   Figure~\ref{fig:band1_image} and discussed in the caption to
   that figure. }
 \label{fig:HeI}
\end{figure}

The extraction of the sky spectrum is described in
\S\ref{sec:DataAnalysis}.  An example sky spectrum is shown in
Figure~\ref{fig:HeI}.  We see that the peak spectral intensity of
the He\,1.0833\,$\mu$m line is $I_\nu\approx 0.23\,{\rm MJy\,ster^{-1}}$,
brighter than the zodiacal background at $0.15\,{\rm MJy\,ster^{-1}}$.
The intrinsic width of this triplet is $<1\,$\AA. The low spectral
resolution of the SPHEREx variable linear filter explains the
noticeable full width at half-maximum of 0.026\,$\mu$m seen in this
figure.

For spectral lines, a better choice for intensity is the line-integrated
photon intensity, $\mathcal{I}=\int I_\nu/(h\nu)d\nu$, expressed
in Rayleigh\footnote{1 Rayleigh or $R=10^6/(4\pi)\, {\rm
phot\,cm^{-2}\,s^{-1}\,ster^{-1}}$ is the traditional unit for
surface brightness in the fields of aeronomy and the  study of
diffuse ionized gas in the Milky Way.}. This quantity  is approximated
by $\int I_\nu d\nu/(h\nu_0)$ where $\nu_0$ is the center frequency
of the line.

As can be gathered from Figure~\ref{fig:HeI}, our data analysis
yields $F\equiv  \int I_\nu d\lambda$ whose units are ${\rm
MJy\,ster^{-1}\,\mu m}$.  Noting that $\vert d\lambda \vert = d\nu
(c/\nu^{2})$ we find $\mathcal{I}=(a\nu_0)/(ch\alpha)F$ with
$a=10^{-17}\times 10^{-4}$ to convert ${\rm MJy\,ster^{-1}}\,\mu$m
to CGS units (${\rm erg\,cm^{-2}\,s^{-1}\,Hz^{-1}\,ster^{-1}}$) and
$\alpha=10^6/(4\pi)$ to convert CGS  to Rayleigh.  Putting all this
together,
 $$
	\mathcal{I}_{\rm 1.0833\,\mu m}=1.751\times 10^4F_{\rm
	1.0833\,\mu m}\,R\ .
 $$
In Figure~\ref{fig:HeI} we see $F_{\rm 1.0833\,\mu m}=0.006\, {\rm
MJy\,ster^{-1}\,\mu m}$. Thus, $\mathcal{I}_{\rm 1.0833\,\mu
m}=105\,R$.  For comparison, the zodiacal light is
$\mathcal{I}/\mathcal{R}=138\,R$ over spectral width of
$\delta\lambda=0.026\,\mu$m.

\section{Metastable Helium}
	\label{sec:MetastableHelium}

It was well known to the early atomic physicists that helium, unlike
hydrogen, exhibited two sets of lines. These were empirically
attributed to ``para-helium'' and ``ortho-helium''. Their origin
was not understandable in the ``old" quantum theory of Bohr \&
Sommerfeld.  In 1926, Heisenberg, with his ``new quantum" theory,
provided the modern explanation, namely that helium has two
spectroscopic families: singlet ($S=0$) and triplet ($S=1$). We now
know that transitions across the spin families are not allowed
(``semi-forbidden") which explains the observed distinct families.

\begin{figure}[hbtp]
 \plotone{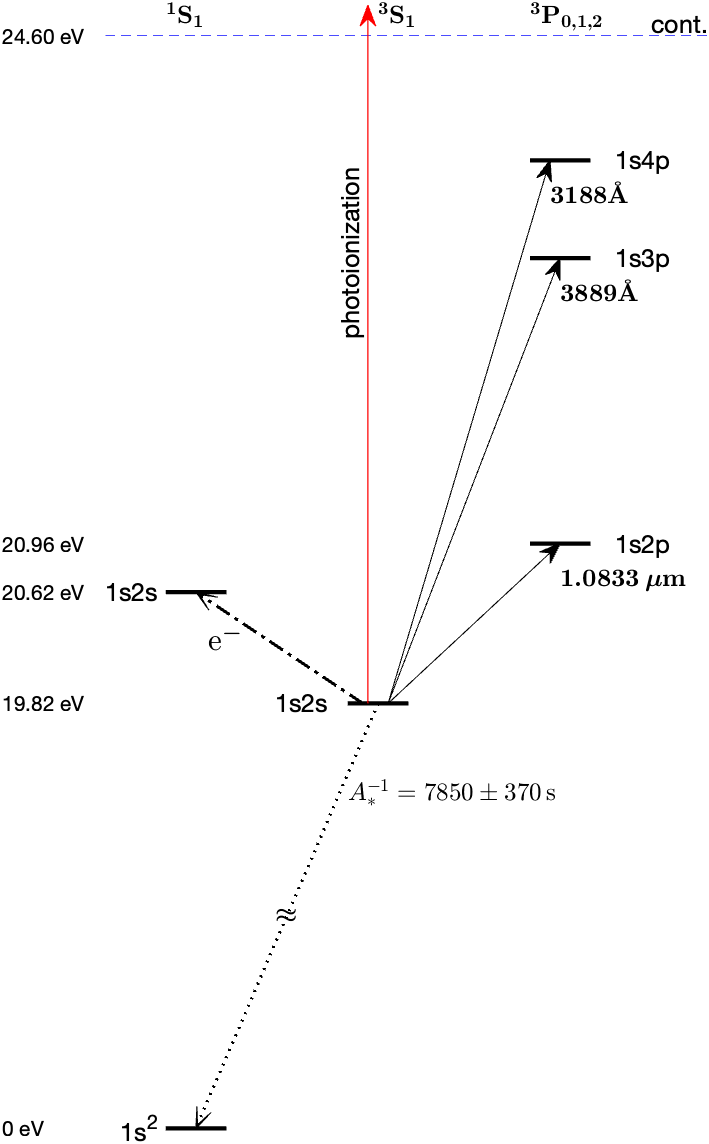}
  \caption{\small Simplified Grotrian diagram for He~I with focus
  on metastable helium.  The energy scale (left side), apart from
  a break (indicated by $\approx$ sign) between the ground state
  and the metastable level, 1s2s\,${\rm ^3S_1}$, is linear.  Key
  allowed transitions of the metastable level are shown by solid
  lines.  The semi-forbidden decay from 1s2s\,${\rm ^3S_1}$ to the
  ground state, with small A-coefficient, is shown by the dotted
  line.  Collisions with electrons can excite atom to 1s2s\,${\rm
  ^1S_1}$ (labeled, e$^-$)  or 1s2p\,${\rm ^3P}$ level (not labeled
  to avoid clutter) and can also de-excite metastable helium to
  ground state (not labeled).  Photons with energy $>4.768\,$eV
  have the capacity to photo-ionize metastable helium (red line). }
 \label{fig:He_Grotrian}
\end{figure}

The lowest level of para-helium is ${\rm 1s^2\,^1S_0}$ (the true
ground state), while that for ortho-helium is ${\rm 1s2s\,^3S_1}$.
A simplified Grotrian diagram is presented in Figure~\ref{fig:He_Grotrian}.
The lifetime of ortho-helium in the lowest state is $A_*^{-1}$ where
$A_*$ is the A-coefficient of the semi-forbidden 1s2s~${\rm
^3S_1\rightarrow {\rm  1s^2}\,^1S_0}$ transition.  The modern value
for $A_*^{-1}$ is about 2.2\,hours, an eternity for permitted
transitions. For this reason, helium in the ${\rm 1s2s\,^3S_1}$
state is called metastable helium.  In atomic physics, metastable
helium is sometimes labeled ${\rm ^4He^*}$.

Metastable helium is long-lived enough to open up all sorts of
interesting phenomena. For example, $^4{\rm He}^*$ is magnetic
\citep{P68}, unlike $^4$He.  Due to the high internal energy ${\rm
^4He^*}$ is used for QED precision tests.  The bosonic nature opens
up novel tests of quantum optics (see \citealt{amp+21}).  Bosonic
${\rm ^4He^*}$ and fermionic ${\rm ^3He^*}$ allow novel studies of
Bose-Fermi mixtures \citep{toh+23}. Finally, it is the metastable
state of both He and Ne that lies at the heart of He-Ne lasers.

Returning to the topic at hand: the He~I\,1.0833$\mu$m line results
from 1s3p~${\rm ^3P}\rightarrow\ {\rm  1s2s~ ^3S_1}$. Historically,
in astronomy, the blue optical triplet $\lambda$3889\,\AA\ (1s4p~${\rm
^3P}\rightarrow\ {\rm 1s2s~ ^3S_1}$) was observed first. These lines
are shown in Figure~\ref{fig:He_Grotrian} and the relevant atomic
data are summarized in Table~\ref{tab:mHe_lines}.

\begin{deluxetable}{lrlr}[hbtp]
\label{tab:mHe_lines}
 \tablecaption{Two major triplets of  metastable He}
 \tablewidth{0pt}
 \tablehead{
 \colhead{$\lambda$\,(nm)}&
 \colhead{$u$} &
 \colhead{$A_{ul}\,({\rm s^{-1}})$} &
 \colhead{$f_{lu}$}
 }
 {\small
 \startdata
1,083.205 7482 & 1s2p ${\rm ^3P^o_0}$ & $1.02\times 10^7$ &0.0599 \\
1,083.321 6761 & \ \ \ \  ${\rm ^3P^o_1}$ & \ \ \ "  & 0.1796\\
1,083.330 6454 & \ \ \ \  ${\rm ^3P^o_2}$ & \ \ \ " & 0.2994\\
\hline
388.970 65567 & 1s3p ${\rm ^3P^o_0}$& $9.47\times10^{6}$ & 0.0072\\
388.974 75124 & \ \ \ \  ${\rm ^3P^o_1}$&	\ \ \ " & 0.0215\\
388.975 08392 & \ \ \  \ ${\rm ^3P^o_2}$&	\ \ \ " & 0.0358\\
\enddata
}
 \tablecomments{\small  $\lambda$ is the wavelength of the transition.
 $u$ is the upper state. For the lines listed here, the lower state
 ($l$)  is 1s2s ${\rm ^3S_1}$.  $A_{ul}$ is the A-coefficient of
 the $ul$ transition and $f_{lu}$ is the corresponding oscillator
 strength.  The sum of the oscillator strengths for the triplet
 1.0833\,$\mu$m line is 0.5389 and 0.0645 for  the 3889\,\AA\
 triplet.  The sum of the oscillator for all the allowed transitions
 from 1s2s~${\rm ^3S_1}$ up to 1s10s is 0.66. The effective wavelength
 for each triplet, calculated by weighting the transition wavelengths
 by $g_u$ (the degeneracy of the upper state), is 1083.3138\,nm and
 388.9745\,nm.   }
\end{deluxetable}

\subsection{A brief history of $A_*$}

Astronomers starting with Viktor Ambartsumian in the thirties were
aware of the possible diagnostic value afforded by metastable helium.
However, in those early days, even the mode of decay (two-photon
versus magnetic dipole) was not known, let alone a rough value for
$A_*$.  \cite{G39} attributed the brightness of the optical lines
of the triplet family relative to those of the singlet family, in
the solar corona, to the ``metastability" of the ${\rm ^3S_1}$
level.  The line became a mainstay of solar research (e.g., 
\citealt{zd63}) and, given its ability to trace the chromosphere
and outflow, was applied to study active stars (e.g., \citealt{Z76}).

\cite{M57}, assuming two photon decay, estimated a value of
$A_*=2\times 10^{-5}\,{\rm s^{-1}}$. \cite{dd68} showed an error
in this calculation and revised the estimate to $<4\times 10^{-9}\,{\rm
s^{-1}}$ \citep{dvd69}.  The small value of $A_*$ raised the exciting
possibility of a detectable population of metastable helium in
astronomical settings. Rees, Sciama \&\ Stobbs (1968) \nocite{rss68}
extended the calculations to include optical absorption  from cluster
gas and the intergalactic medium\footnote{The article ends with the
statement that 1-photon decay calculations were on the verge of
being carried out and this channel was likely to be faster than the
2-photon decay channel.} while \citet{S68} called for observations
of Galactic ISM in the 1.0833\,nm line.

By 1970s QED calculations became possible and it became immediately
clear that the magnetic dipole (M1) interaction dominates over two-photon
decay (cf. \citealt{D71}).  Experimental measurements carried out
two years later found $A_*\approx 2.3\times 10^{-4}\,{\rm s^{-1}}$
with an error of a third of the value \citep{mw73}. Currently, the
best experimental measurement value is $A_*=(1.274\pm 0.06)\times
10^{-4}\,{\rm s^{-1}}$ which corresponds to $A_*^{-1}$  of  $7850\pm
370\,$s (\citealt{hdb+09a}; see this article for a history of QED
calculations of $A_*$).

\subsection{Line(s) of metastable helium: Astronomy}

\begin{figure*}[hbt!]   
 \centering
  \includegraphics[width=0.7\textwidth]{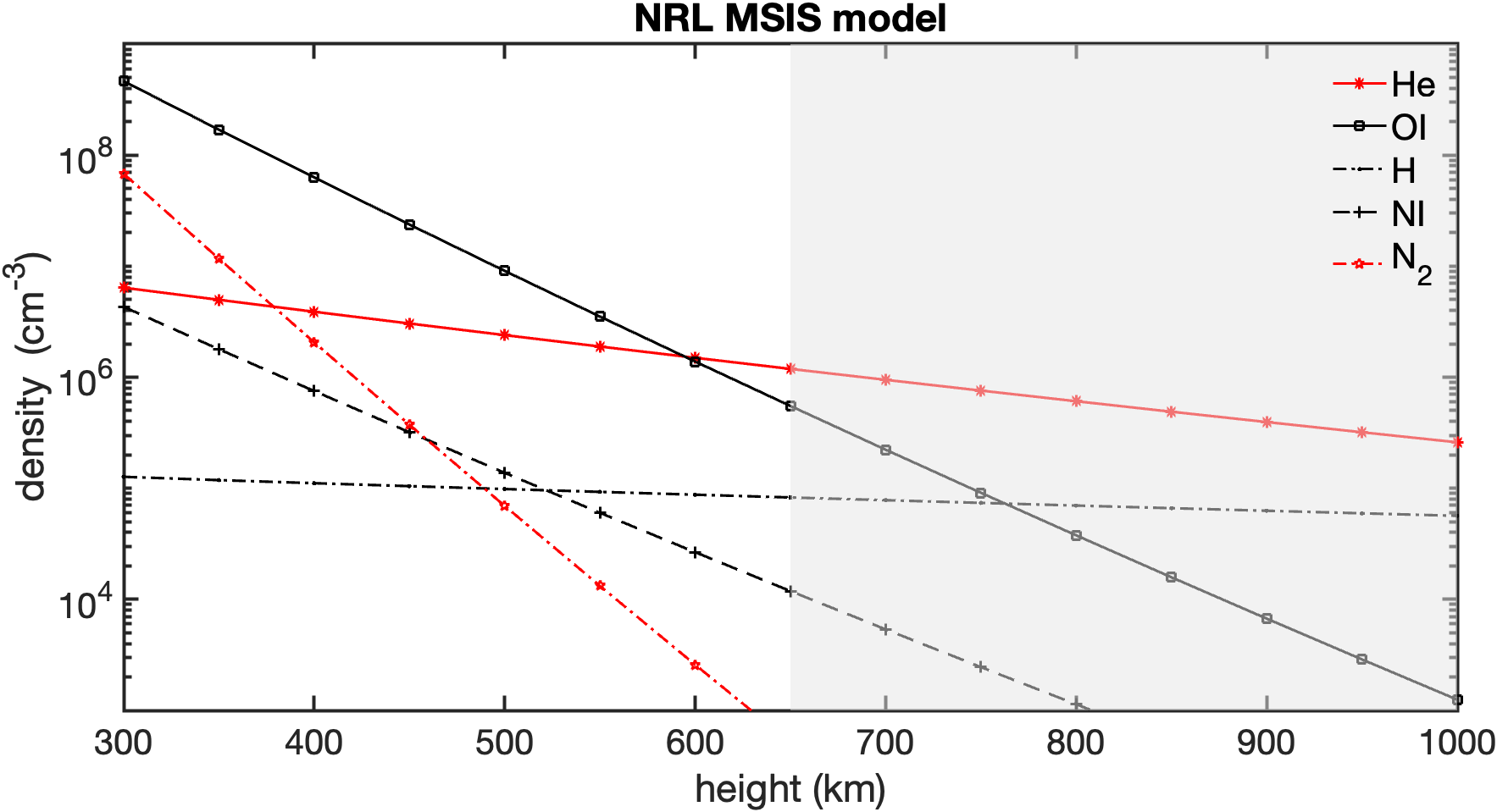}
   \caption{\small Naval Research Laboratory (NRL)  Mass Spectrometer
   and Incoherent Scatter radar (MSIS) model of the neutral species
   in the upper atmosphere  for a location defined by terrestrial
   longitude of $\lambda=90^\circ$ and terrestrial latitude of
   $\phi=30^\circ$ on UT 2025 June 20 at 12:00 am.  Throughout the
   thermosphere, the temperature of the neutral species is about
   1100\,K.  The atmosphere above the orbit of SPHEREx (height,
   650\,km) is shown in gray.  The  model data were generated using
   NRLMSIS 2.0 (\url{https://ccmc.gsfc.nasa.gov/models/NRLMSIS~2.0/}).}
  \label{fig:NRLMSIS}
\end{figure*}

The first detection of metastable helium\footnote{Absorption of the
${\rm  1s2s\ ^3S_1\rightarrow 1s4p\ ^3P_{0,1,2}}\,\lambda 3888$\,\AA\
triplet} seems to be by \cite{A49}.  \citet{ihh+09} and \cite{gk12}
revisited the same topic (Galactic ISM) but with the He~I 1.0833\,$\mu$m
line instead. As noted by solar astronomers, lines of metastable
helium, can trace the transition from outflow to winds \citep{uw81,dsl92}
and, as such, has been pressed in the study of young stars with
inflows and outflows (e.g. \citealt{efk+03}), of giant stars stars
\citep{dss09} and R Corona Borealis stars \citep{cgz13}. The line
has also been used in the study of explosive and / or energetic
phenomena: galactic nuclei \citep{ak69}, classical novae \citep{nta13},
and tidal disruption events \citep{hcr+19}.  In exoplanets,
investigations of He~I 1.0833\,$\mu$m began in 2018 \citep{oh18,sse+18}.
This diagnostic has now transitioned from a cottage industry to an
industrial operation.

\subsection{Line(s) of metastable helium: Aeronomy}

During the day, the He~1.0833\,$\mu$m line is as bright as 15\,kilo
Rayleigh (kR); see, for example, \citet{hc69}. In the terrestrial
context, it was first discovered by observations taken at Moscow
in 1957, following a particularly strong auroral
storm\footnote{Coincidentally, the International Geophysical Year,
IGY, began in July 1957.} (see \citealt{S64} and the references
therein).  However, most of the ground-based observations were made
after the sun set or before the sun rose to avoid Rayleigh scattering
from the sun. Once the thermosphere is no longer illuminated by
sunlight, the line disappears and so the usable time is restricted
to SZA,  less than, say, 130$^\circ$; here SZA is the zenith angle
of the sun (wth 90$^\circ$ corresponding to sunset or sunrise).
The underlying physics for the production of metastable helium in
the thermosphere was developed and clarified by \cite{bl93}.  The
further discussion of this topic is postponed to \S\ref{sec:Thermosphere}.

\section{Helium in the atmosphere}
	\label{sec:Thermosphere}

At sea level, in the dry atmosphere, helium, with an abundance of
5.24 ppm (parts per million), is the sixth most abundant species
after N, O, Ar, CO$_2$ and Ne.  This is followed by Kr, SO$_2$ and
CH$_4$ ($>1\,$ppm).  Other trace species such as H$_2$, Xe, O$_3$,
NO$_2$, I$_2$ and CO and NH$_3$ have abundances below 1\,ppm.  The
low concentration of helium makes commercial extraction of helium
from the air impractical.  Instead, it is obtained as a by-product
of the extraction of natural gas.

The working hypothesis is that helium in the atmosphere originates
from the decay of radioactive elements (particularly, uranium and
thorium).  The outgassing of helium from the crust is estimated to
be $2\times 10^6\,{\rm atom\,cm^{-2}\,s^{-1}}$ \citep{M63}.  In
recent times, as a consequence of the increased extraction of natural
gas, the concentration of both $^4{\rm He}$ (and, surprisingly,
${\rm ^3He}$) has been gradually increasing \citep{bsp+22}.

For the astronomers, I provide a short summary of the vertical
structure of the atmosphere.  The lowest level is the troposphere
[0--10\,km] in which, as we know from experience, the temperature
falls with height; the stratosphere [10--50\,km] where, due to
heating provided by solar photo-dissociation of ozone, the temperature
gradient is reversed; the mesosphere [50--80\,km] where the temperature
gradient becomes negative again; the thermosphere [80-600\,km] where
the temperature gradient becomes positive again due to heating
provided by solar photo-dissociation of O$_2$; and finally the
exosphere, which depending on solar activity, variously starts at
600\,km to 1,000\,km.

The run of dominant neutral species in the altitude range of interest
for SPHEREx is shown in Figure~\ref{fig:NRLMSIS}. It is interesting
to note that the dominance of helium at the top edge of the
thermosphere was first inferred by orbital decay  of the Echo~1
``balloon-satellite" \citep{N61}.  For neutral species, the highest
temperatures in the thermosphere range from 1,000\,K to 2,000\,K
(depending on solar activity). Above about 1,000\,km, the atmosphere
becomes a gentle wind that carries the lightest elements (H, He)-- the
exosphere.

The loss of helium from the earth's atmosphere is an important
problem in Earth science.  \cite{M63} and \cite{K73}, although
dated, are good starting points while \cite{H82} provides an overview
of outgassing of the planets (including earth) of the solar system.
Briefly, the earth loses hydrogen and helium as a result of three
processes. These are {\it (i)} Jean's escape (fast moving particles
on the Maxwellian tail escape the earth), {\it (ii)} charge exchange
of an atmospheric helium atom with energetic solar protons, and
{\it(iii)} polar wind escape (acceleration of ions along open field
lines at the magnetic poles).  Currently, it is estimated that the
earth loses 3\,kg\,s$^{-1}$ of hydrogen and 50\,g\,s$^{-1}$ of
helium \citep{cz09}.  All three processes discussed above are
sensitive to the activity of the sun.  The orbital longevity of low
earth orbit (LEO) satellites is sensitive to the expansion and
contraction of the thermosphere.

\section{Metastable helium in the thermosphere}
	\label{sec:MetastableHeliumThermosphere}
	
In aeronomy, it appears that \cite{bl93} is considered the defining
document of all relevant theoretical processes with respect to the
intensity of the He~I\,1.0833\,$\mu$m line. \cite{wkg+05} present
an analysis of a comprehensive (radar and Fabry-P\'erot) year-long
program at the Arecibo Observatory aimed at studying metastable
helium.

\begin{figure}[htbp]    
 \plotone{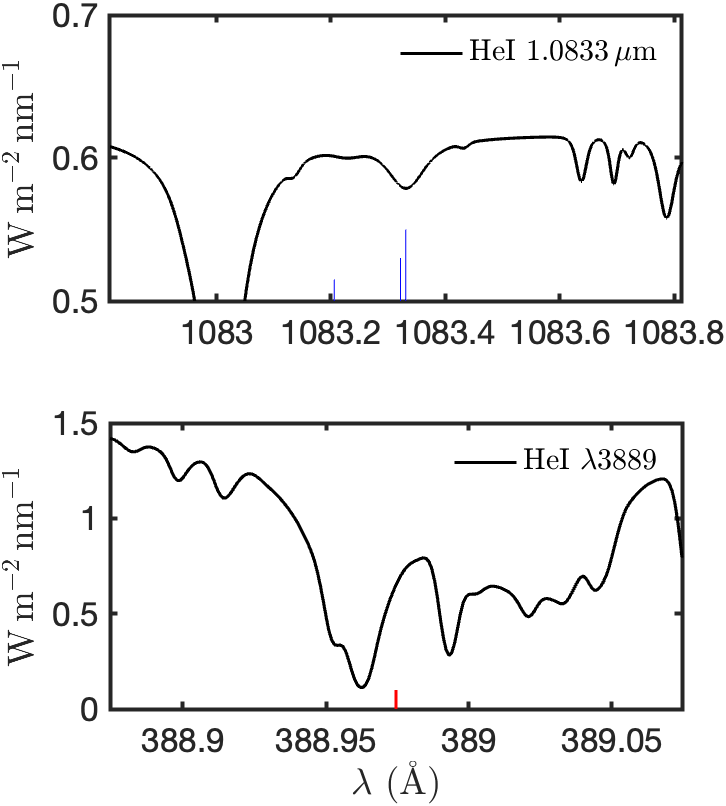}
  \caption{\small Zoom-in of the TSIS-1 HSRS solar spectrum centered
  on wavelengths of  He~I 1.0833\,$\mu$m and He~I\,$\lambda 3889$.
  For the 1.0833\,$\mu$ line, vertical blue stubs mark the three
  lines with height proportional to strength. For $\lambda 3889$
  the red vertical stub marks the effective wavelength of the triplet
  (see notes to Table~\ref{tab:mHe_lines}).  The data were obtained
  from \url{https://lasp.colorado.edu/lisird/data/tsis1_hsrs_p1nm};
  see \cite{crh+23} for further details.}
 \label{fig:Solar_HeI}
\end{figure}

In the thermosphere, a number of processes can produce metastable
helium: collisional excitation with energetic electrons, recombinations
with low energy electrons and (successive) charge exchange of fast
moving alpha particles in the solar wind plowing into the
thermosphere (see, for example, \citealt{R73}).  However, the
dominant process is collisional excitation by freshly minted
photoelectrons (see \citealt{bl93}). Such energetic electrons are
produced by molecules and/or atoms that absorb solar EUV photons.

The destructive pathways for metastable helium, other than radiative
decay to the ground state, are as follows: photoionization, electron
collisional de-excitation to the ground state,  and electron collisional
excitations to singlet levels, especially adjacent energy levels
(e.g., ${\rm 1s2s\,^1S_0}$, 0.79\,eV above 1s2s~${\rm ^3S_1}$; ${\rm
1s2p\,^1P_0^{\rm o}}$, 1.4 eV above) which then undergo rapid
radiative decay.  These processes are illustrated in
Figure~\ref{fig:He_Grotrian}.  In the upper thermosphere, 
photoionization by sunlight is  important whereas below 350\,km, Penning
ionization\footnote{Collisions of the sort He$^*$+X where X is
another species result in ionization of X. In this context, X is
O$_2$, N$_2$, O$^0$ and H$^0$. This process, discovered by F.\ M.\ Penning
of Philips Natuurkundig Laboratorium, Eindhoven, Netherlands, is
the key mechanism that underlies the operation of house-hold tube
lights.} is very effective in suppressing metastable helium.

Here, we treat the column density of metastable helium above the
orbit of SPHEREx as a free parameter whose value would be determined
from the SPHEREx observations. The proposed model for the origin
of the He~I line is simple: the source of 1.0833\,$\mu$m emission
is {\it primarily} due to resonance scattering of solar photons by
metastable helium.

\section{Resonance Scattering of Solar Photons}
	\label{sec:ResonanceScattering}

Our goal is to compute the absorption cross section for the lines
of interest. We start the exercise by noting that the thermal
broadening of the helium atoms, $\sigma_v\approx 1.4T^{1/2}_3\,{\rm
km\,s^{-1}}$, is small, as are the wind velocities in the thermosphere;
here, $T=10^3T_3\,$K is temperature of the helium gas. For the ``top
of the atmosphere" solar spectrum, we elected to use the ``Total
Spectral Solar Irradiance Sensor-1 Hybrid Solar Reference Spectrum"
(TSIS-1 HSRS; \citealt{crh+21,crh+23}) because of its higher spectral
resolution (0.01\,nm) relative to SOLSPEC (which ranges from 0.6
to 9.5\,nm; \citealt{mdb+18}).

The solar spectrum in the vicinity of the 1.0833\,$\mu$m line and
the $\lambda 3889$ line is shown in Figure~\ref{fig:Solar_HeI}.
The spectral flux density at 1.0833\,$\mu$m is $f_\lambda=0.58\,{\rm
W\,m^{-2}\,s^{-1}\,nm^{-1}}$ which translates to $2.27\times
10^{-9}\,{\rm erg\,cm^{-2}\,s^{-1}\,Hz^{-1}}$ or $N_\nu(0)=1.24\times
10^3\,{\rm phot\,cm^{-2}\,s^{-1}\,Hz^{-1}}$.

The absorption cross-section is
 $$
	\sigma_{lu}(\nu) = \frac{\pi e^2}{m_e c}f_{lu}\phi_\nu
 $$
where $f_{lu}$ is the oscillator strength (see Table~\ref{tab:mHe_lines}),
$\phi(\nu)d\nu$ is the probability of photon being observed over
the interval, $[\nu,\nu+d\nu]$ and   $\int \phi_\nu d\nu=1$.  The
optical depths are small, so we can ignore the damping wings.
Metastable helium can be excited to any of the three levels of ${\rm
^3P_{0,1,2}}$.  At the spectral resolution of relevance to this
paper the resulting lines are at the same wavelength.  So, what
matters is the sum of the three oscillator strengths.  The scattering
rate per atom is  simply $\mathcal{R}_{1.0833}=(\pi e^2/m_ec) f_{{\rm
^3S_1\rightarrow{} ^3P_{0,1,2}}}N_\nu(0)$
 $ = 17.7\,{\rm phot\,atom\,s^{-1}}$.

The photon intensity of the scattered 1.0833\,$\mu$m line is given
by $n_{\rm He^*}L \mathcal{R}_{1.0833}/(4\pi)$ where $n_{\rm He^*}$
is the number density of metastable helium and $L$ is the length
of the column of metastable helium. Normalizing to a path length
of 100\,km we find
 \begin{equation}
	\mathcal{I}_{1.0833} = 177 n_{\rm He^*}L_2 \ R\ .
		\label{eq:Prediction_1083}
 \end{equation}
where $L_2=L/({\rm 100\,km})$ and $n_{\rm He^*}$ is in cm$^{-3}$.

A similar exercise can be carried out for the 3889\,\AA\ triplet
(Table~\ref{tab:mHe_lines}). The spectral flux density at this
wavelength is ${\rm 0.56\,W\,m^{-2}\,nm^{-1}}$.  The corresponding
scattered-photon intensity is
 $$
	\mathcal{I}_{3889}=0.95n_{\rm He^*}L_2\,R \ .
		\label{eq:Prediction_3889}
 $$

\subsection{Expectations}

In this subsection, we confront our model expectations with existing
data and accepted models for metastable helium in the thermosphere.

A recent development in aeronomy has been the use of lidar: shining
a bright beam at a wavelength of 1.0833\,$\mu$m into the thermosphere.
Currently, there are two facilities: Deutsches Zentrum f\"ur Luft-
und Raumfahrt (DLR; German Aerospace Center; Munich; $48^\circ$\,N,
11$^\circ$E; \citealt{kgb+22}) and the Chinese Meridian Project
group operating from Danzhou, Hainan Province, China ($19.5^\circ$N,
109$^\circ$E; \citealt{zlx+24}). The DLR group conducted measurements
during the northern winter period, when the intensity of
He~I\,1.0833\,$\mu$m is expected to be the brightest \citep{gk25}.
The inferred densities for metastable helium were a few atom
cm$^{-3}$.  Putting this density in Equation~\ref{eq:Prediction_1083}
we find intensities approaching 1\,k$R$ that match the highs seen
in Figure~\ref{fig:HeI_2025_06_20}.  Separately, if the typical
density of metastable helium atoms is 1\,cm$^{-3}$, then from
Figure~\ref{fig:NRLMSIS}, we see that the fractional abundance of
metastable helium is very small $\lesssim 10^{-6}$.

\subsection{Photoionization}

In the upper thermosphere, the primary process for the destruction
of metastable helium is photoionization \citep{bl93}.  The ionization
potential of metastable helium is IP=4.768\,eV (see
Figure~\ref{fig:He_Grotrian}) corresponding to $\lambda=2600\,$\AA.
Incidentally, recently, $\nu_{\rm ip}={\rm IP}/h$ has been measured
with high precision, $1,152,842,742.7082(55)\,$MHz \citep{cga+25}.
This value appears to be in discrepancy with the best QED calculations
\citep{pyp21} by $9\sigma$.

The photo-ionization rate per metastable atom is $\mathcal{R}_{\rm
pi} =\int_{\nu_{\rm pi}}^\infty f_\nu/(h\nu)\sigma_{\rm pi}(\nu)d\nu$,
where $\sigma_{\rm pi}(\nu)$ is the bound-free cross-section at
frequency, $\nu$, and $f_\nu=\lambda f_\lambda/\nu$ is the solar
spectral flux density.  For this calculation high spectral resolution
is not needed. In view of this, for the solar
intensity spectrum, we elect to use the SOLSPEC spectrum
\citep{mdb+18}.


In \S\ref{sec:Photoionization} we summarize the history of improvements
in the calculation of $\sigma_{\rm pi}$.  There we noted that the
experimental data lie above the calculation of \cite{N71} and below
that of \cite{dsb+20}.  The corresponding values for $\mathcal{R}_{\rm
pi}$ are $1.4\times 10^{-3}\,{\rm s^{-1}}$ and $2.7\times 10^{-3}\,{\rm
s^{-1}}$, respectively. We adopt the average, $\mathcal{R}_{\rm
pi}\approx 2\times 10^{-3}\,{\rm s^{-1}}$, in excellent agreement
with \cite{bl93}.

\subsection{Other processes}

\begin{figure}[htbp] 	
 \plotone{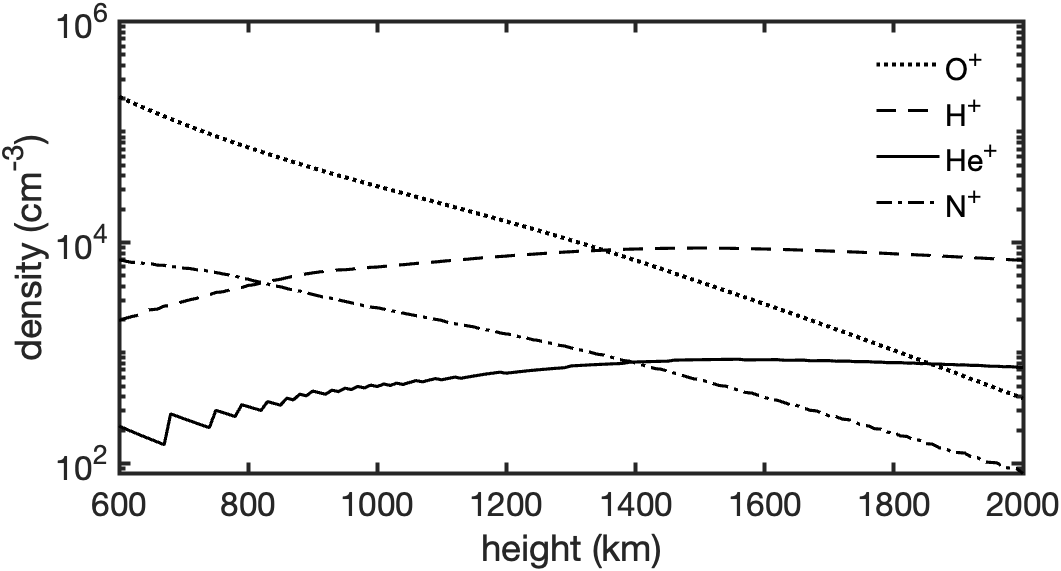}
  \caption{\small Run of densities of dominant ions with height at
  terrestrial coordinates: $\phi=30^\circ$, $\lambda=90^\circ$ at
  UT 2025-June-20, 12 noon. The data were obtained by running
  IRI-2020 (\url{https://kauai.ccmc.gsfc.nasa.gov/instantrun/iri/})
  hosted by International Reference Ionosphere
  (\url{https://irimodel.org/}).} 
 \label{fig:IRI_Hep}
\end{figure}

\begin{figure}[htbp]    
 \plotone{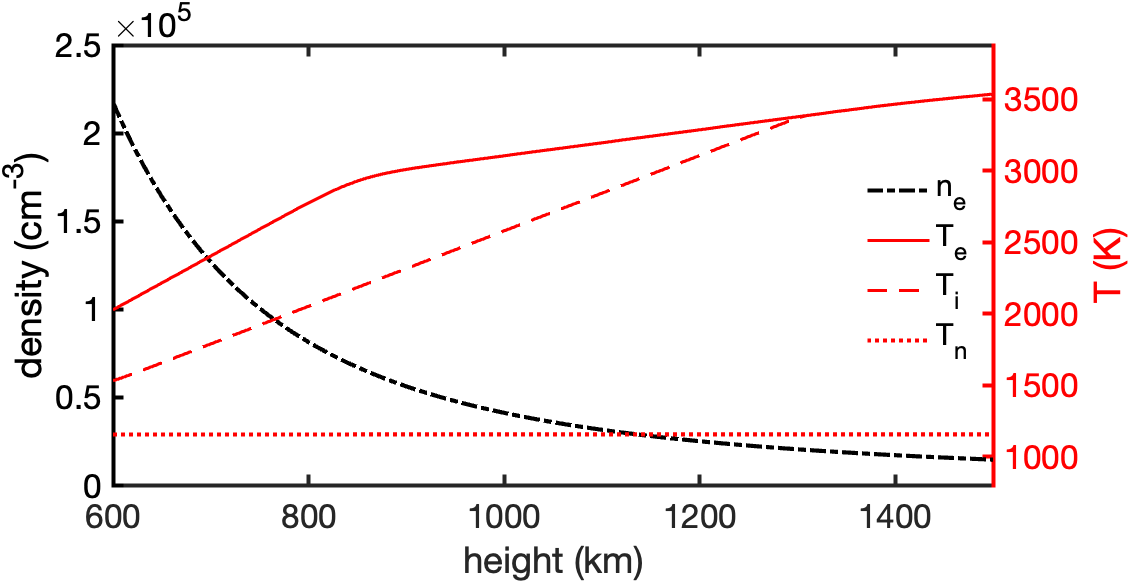}
  \caption{\small The run of  electron density and temperature(s)
  with height.  See caption to Figure~\ref{fig:IRI_Hep} for details
  of the model.} 
 \label{fig:IRI_ne} 
\end{figure}

For completeness, we now investigate other processes that involve
metastable helium.  In order not to distract the reader from the
primary focus of the paper, the background of the secondary processes
is summarized in \S\ref{sec:OtherProcesses}.  The run of ion densities
is shown in Figure~\ref{fig:IRI_Hep} while the run of electron
density and temperature(s) can be found in Figure~\ref{fig:IRI_ne}.
The (astronomical) reader may be surprised to learn that the
ionosphere is mildly ionized, with ionization fraction typically
less than 1\%. Furthermore, as is clear from Figure~\ref{fig:IRI_ne},
each species (neutrals, ions, protons, ions) have their own
temperatures -- a situation not routinely encountered in studies of the
interstellar medium.

First, we investigate the production of metastable helium by
recombination.  Three quarters of the recombinations will proceed via
the triplet ladder, with all ending in 1s2s~${\rm ^3S_1}$. In other words,
three-quarters of the recombinations result in metastable helium.  We
adopt an electron density, $n_e=10^5\,{\rm cm^{-3}}$ and an electron
temperature of 3,000\,K (Figure~\ref{fig:IRI_ne}).  Applying
Equation~\ref{eq:Recombination} we find the recombination timescale
to be $t_r=(\sfrac{3}{4}n_{e}\alpha)^{-1}\approx1.6\times 10^7\,$s.

Given Figure~\ref{fig:NRLMSIS} we restrict the discussion of
Penning ionization
of hydrogen atoms by metastable helium.
For this endothermic
reaction, the rate coefficient, over a wide temperature range,
is $Q=5\times 10^{-10}\,{\rm cm^3\,s^{-1}}$ (\S\ref{sec:OtherProcesses}).
From Figure~\ref{fig:NRLMSIS} we see that $n_{\rm H}\approx 10^5\,
{\rm cm^{-3}}$, over a wide range of altitude. The timescale for
the reaction is $(n_{\rm H}Q)^{-1}\approx 2\times 10^4\,{\rm s}$,
much slower than that for photoionization. The density of neutrals
increases with decreasing height, whereas solar photoionization
remains the same. This calculation shows why at lower heights Penning
ionization limits the density of metastable helium atoms.

We finally consider collisions with electrons.  The rate coefficient
for excitation to the 1s2s~${\rm ^3P}$ state is $q=0.46\times
10^{-8}\, {\rm cm^3\,s^{-1}}$ while  $q=0.76\times 10^{-8}\,{\rm
cm^3\,s^{-1}}$ for 1s2s~${\rm ^1S_1}$ is and $q=0.23\times 10^{-8}\,{\rm
cm^3\,s^{-1}}$ for de-excitation to the ground state.  Given
$n_e\approx 10^5\,{\rm cm^{-3}}$, these timescales are of the order
of $10^3\,$s,  a factor of two longer than the photo-ionization
timescale.  
Separately, note that the production of 1.0833\,$\mu$m photons by electron
collisional excitation simply does not compete with resonance
scattering of solar photons.

\section{Orbital Geometry}
	\label{sec:OrbitalGeometry}
	
In the framework adopted here, the brightness of the He~I line is
proportional to the product of the intensity of the incident sunlight
and the column density of metastable helium atoms. Changes in either
quantity will affect the brightness of the line.  Thus, it is
important to have a complete understanding of the geometry of the
situation before interpreting the data.

\begin{figure}[htbp]   
 \plotone{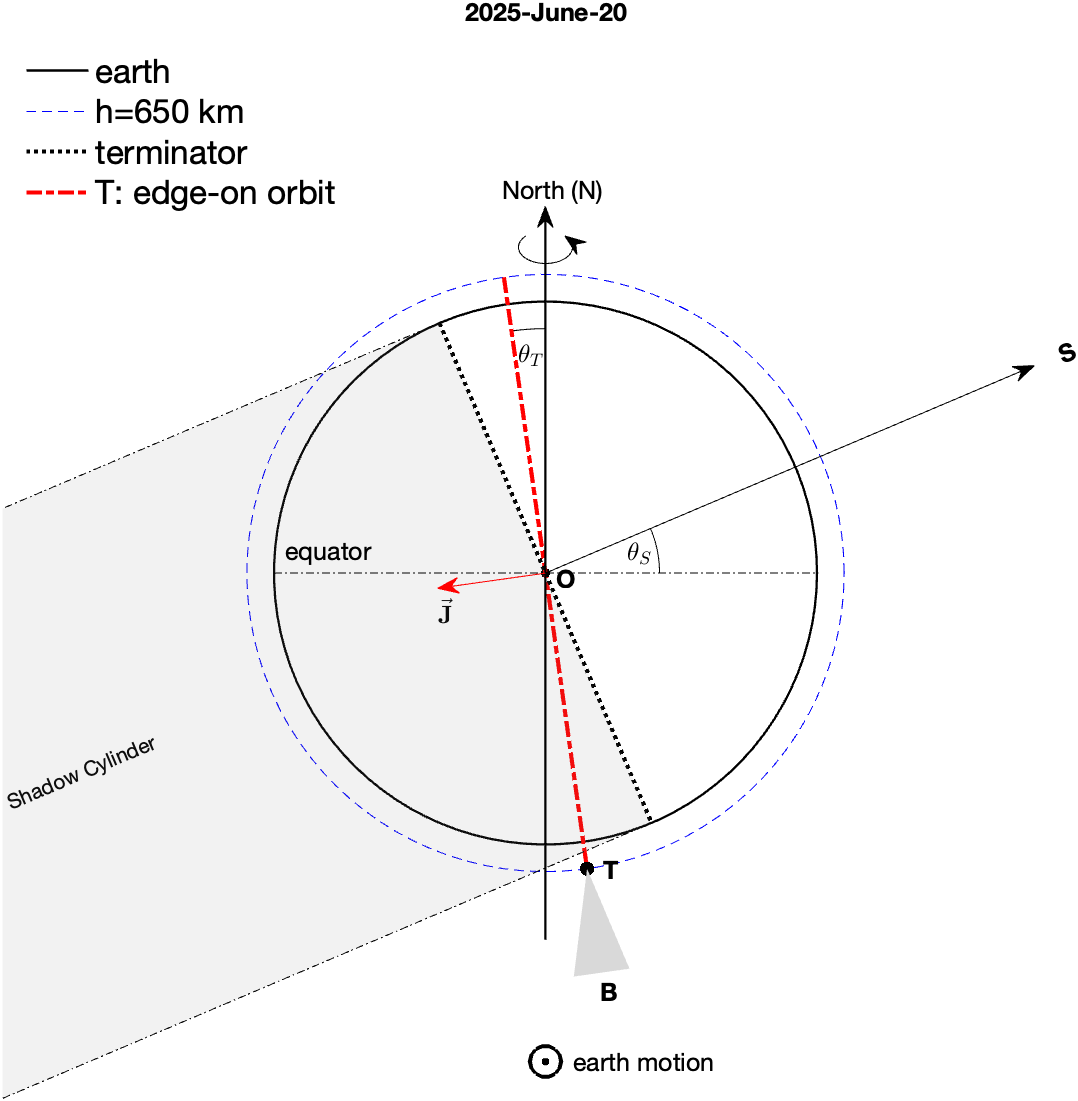} \smallskip
  \caption{\small The edge-on orbit of SPHEREx (tilted dash-dotted
  red  line) at the northern solstice (20th June, 2025).  SPHEREx
  is labeled by ``T".  The  large black circle represents the earth
  with the celestial north pole marked and the sense of earth's
  rotation clearly indicated. The dashed blue curve marks a circle
  whose height above the earth's surface is 650\,km.  The center
  of the sun (``S") and the center of earth (``O") lie  in the plane
  of the paper.  Our view is one of  looking back at earth from
  autumnal equinox (first point of Libra).  [In regard to the symbol
  $\odot$ found at the bottom of the figure: we follow the convention
  of magnetostatics and use the symbol $\odot$ to indicate a vector
  coming out of the paper. Here, the vector under consideration is
  the velocity of earth around the sun.] The sun bears an angle,
  $\theta_S\approx 23^\circ$ with respect to the equatorial plane.
  The orbit of SPHEREx is tilted with respect to the celestial north
  pole, $\theta_T\approx 8^\circ$.  The angular momentum vector of
  the orbit of SPHEREx is marked by a thin red arrow (and labeled
  as $\vec{\mathbf{J}}$). Notice that the east-west velocity component
  of SPHEREx is retrograde with respect to the rotation of earth.
  There is  an additional angle -- the direction of the bore-sight
  (B) with respect to geocentric radial vector of the orbit of the
  satellite.  This angle is under the control of the project.  For
  illustrative purpose, the range in the choice of this angle  is
  shown by a cone with a half-opening angle of 15$^\circ$.  In the
  text, we refer to two angles: SOT and SOB. As can be gathered
  from the labels in this figure, these angles are defined by
  sun-earth(center)-satellite and sun-earth(center)-boresight.}
 \label{fig:SPHEREx_Geometry_N}
\end{figure}

The orbit of SPHEREx is almost a circle with a planned semi-major
axis of 7037\,km and an inclination\footnote{The inclination of a
satellite orbiting the earth is the angle between orbital plane of
the satellite and earth's equatorial plane. A satellite orbiting
in the equatorial plane with motion in the same direction of earth's
rotation (prograde) has an inclination of 0$^\circ$.  A satellite
passing over both poles has an inclination of 90$^\circ$. A satellite
in an equatorial orbit but with motion opposed to earth's motion
(retrograde) has an inclination of 180$^\circ$.} of 97.95$^\circ$.
In Figure~\ref{fig:SPHEREx_Geometry_N}, I present the geometry of
the orbit of SPHEREx relative to earth and the sun at the northern
solstice (UT 2025, June 20).

There are four reference frames: the geocentric ecliptic coordinate
system (longitude:  $\lambda$, latitude: $\beta$), the celestial
(equatorial) frame (right ascension: $\alpha$ and declination:
$\delta$), a frame defined by the magnetic field of the earth and
the geographical frame (longitude: $\lambda$, latitude: $\phi$).
Astronomers prefer $\alpha$ and $\delta$. The most natural coordinate
system for SPHEREx is the geocentric ecliptic frame because in this
frame the orbital orientation is fixed with respect to the sun.
Atmospheric scientists naturally prefer the geographical frame.
For charged particles, the most natural frame is the terrestrial
geomagnetic frame!

The reader has no choice but to skip around the frames as and when
it makes sense. In this paper, I will be using $\lambda$ and $\beta$
because those values are included in the FITS header. As and when
needed I compute quantities such as the position of the sun and the
sun-earth(center)-satellite, traditionally called the SOT angle
(defined and explained in Figure~\ref{fig:SPHEREx_Geometry_N}) and
also the sun-earth(center)-bore-sight (``SOB") angle.  For example,
for the observation presented in Figure~\ref{fig:HeI}, SOT=76$^\circ$
whilst SOB=97$^\circ$.  Finally, unless otherwise stated, the term
``pole" or ``polar" regions refers to the geographical pole.

\section{Variations in the He~I line}
	\label{sec:Variations}

\begin{figure*}[hbtp!]  
 \centering
  \includegraphics[width=0.9\textwidth]{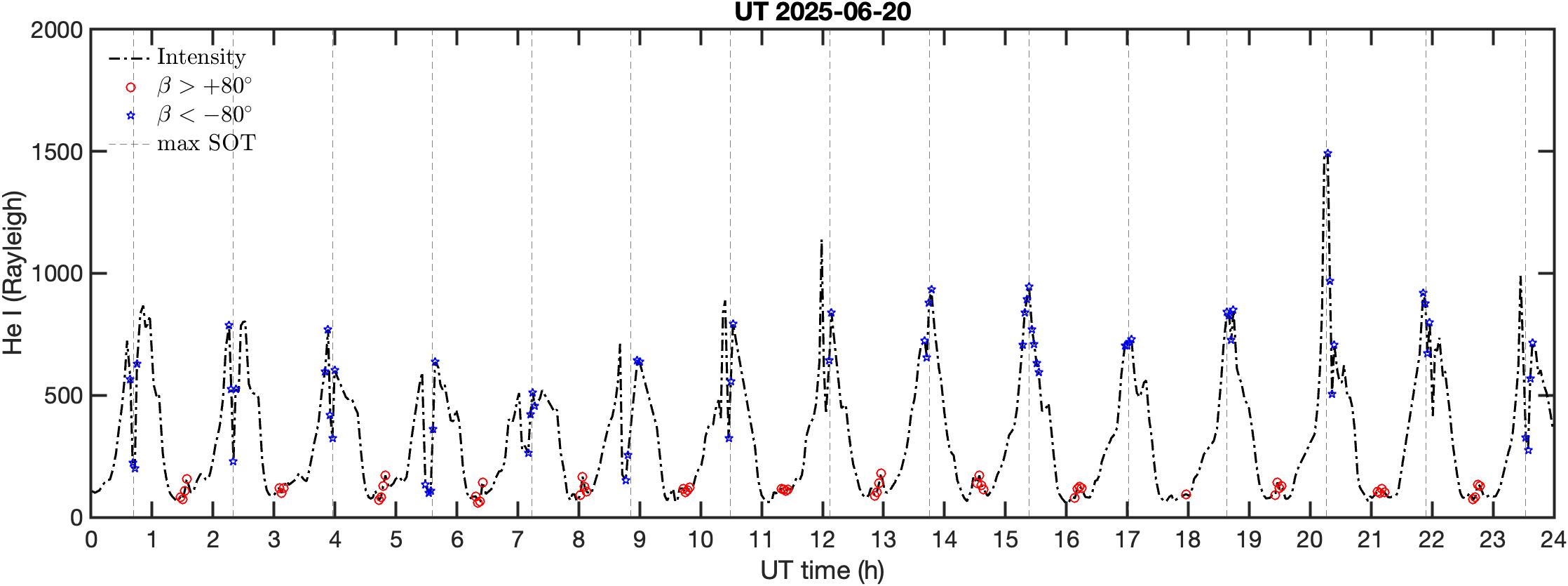}
 \includegraphics[width=0.9\textwidth]{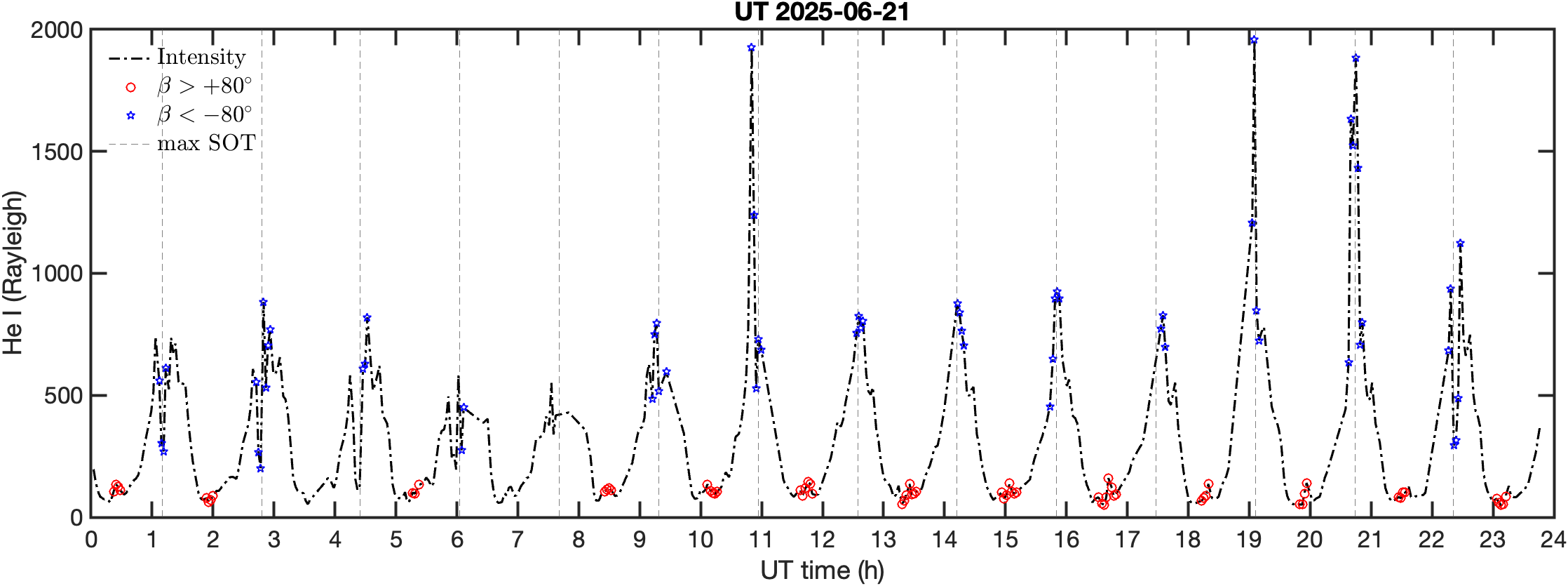}
   \caption{\small The strength of the He~I\,1.0833\,$\mu$m line
   (in Rayleigh, $R$) over the course of two successive days (UT
   date noted in the title). The median uncertainty for the
   measurements is $4.4\,R$.  Note that the excursions to high
   values increased significantly on the second day.  Observations
   at North and South Ecliptic poles ($\vert \beta\vert>80^\circ$)
   are noted in the legend. The vertical lines mark time of maximum
   sun-observer-satellite angle (SOT). }
 \label{fig:HeI_2025_06_20}
\end{figure*}

About every two minutes SPHEREx produces images in each of the six
bands.  I have analyzed band~1 images for two full days that overlap
the summer solstice (2025 June 20 and 21).  The resulting line-integrated
photon intensity of the He\,1.0833\,$\mu$m for these two days is
shown in Figure~\ref{fig:HeI_2025_06_20}. Clearly, the 1.0833\,$\mu$m
emission is periodic. The orbital period of the satellite clearly
emerges from a Lomb-Scargle analysis of the light curve
(\S\ref{sec:Lomb-Scargle}).

As noted in the caption of Figure~\ref{fig:HeI}, the geocentric
ecliptic latitude ($\lambda$) and longitude ($\beta$) of the
bore-sight direction are included in the FITS header.  From
Figure~\ref{fig:HeI_2025_06_20} we see that the He line is strongest
at the South Ecliptic Pole (SEP) and weakest when the satellite is
pointing towards the region of the North Ecliptic Pole (NEP).  

We expect the SOT angle to vary between $90+\theta_S-\theta_T=105^\circ$
(which will happen when the satellite is closest to SEP) and
$90-\theta_S+\theta_T=75^\circ$ (which will happen when the satellite
is closest to the NEP).  I used JPL's Horizons
system\footnote{\url{https://ssd.jpl.nasa.gov/horizons/}.  This
facility provides ephemerides for planets, small bodies, and
satellites.} to compute SOT for SPHEREx.  For June 20, 2025, I find
that SOT varies from 75$^\circ$ to $105^\circ$.  The times of maximum
SOT are marked in Figure~\ref{fig:HeI_2025_06_20}.  The match between
maximum SOT and peak He~I emission means that the emission is the
highest in the vicinity of the south polar region.

As can be concluded from Figure~\ref{fig:SPHEREx_Geometry_N}, the
intensity of the solar light at SPHEREx remains unchanged during
its orbit. Furthermore, recall that the SPHEREx observes close to
the terminator plane. Thus, {\it we are forced to conclude that,
at the northern solstice, the column density of metastable helium
in the upper thermosphere is much higher in the south pole region
relative to the north pole and the equatorial regions.}

\subsection{The winter helium bulge}
	\label{sec:WinterBulge}
	
Persuaded by the striking trend seen in Figure~\ref{fig:HeI_2025_06_20},
namely an increased column density of metastable helium at the south
pole (during northern summer), I perused the aeronomy literature
and found considerable literature on the ``winter helium bulge" --
a notable increase of the concentration of helium in the winter
hemisphere of the thermosphere relative to the summer hemisphere.
This effect was  discovered in 1968 by  by G.\ Keating and E.\ Prior
using satellite drag measurements.

Not only the concentration of helium increases in the wintry pole
but also the supply of photoelectrons generated at the conjugate
point, the bright summer pole, and transported to the wintry pole
along magnetic field lines. As a result, the concentration of
metastable is increased by an order of magnitude.  \cite{stw+15}
present a comprehensive model for the global distribution of helium
at key junctures (solstices and equinoxes).  Figure~2 of this paper
is particularly informative, {\it We predict that SPHEREx will see
strong 1.0833\,$\mu$m emission at the north polar cap during the
forthcoming northern winter season (December 2025).}

\subsection{Solar Activity}
	\label{sec:SolarActivity}
	
Separately, the helium bulge phenomenon is sensitive to the activity
of the sun.  During solar minimum, the thermosphere is cooler
(relative to that at solar maximum). As a result, the helium bulge
is thicker at solar minimum relative to solar maximum.  In addition,
the metastable fraction is directly related to solar activity
(through an increase in the flux of energetic electrons, protons,
and alpha particles, and an increase in the UV flux). Note that
fast-moving solar alpha particles undergo successive charge exchange
as they are slowed down by the neutrals in the upper atmosphere
\citep{R73}.

Bearing this discussion in mind, I reviewed solar activity for June
20 and June 21, 2025. To start with,  I note that the sun reached
solar maximum late last
year\footnote{\url{https://www.swpc.noaa.gov/products/solar-cycle-progression}}.
The average F10.7cm radio
flux\footnote{\url{https://www.swpc.noaa.gov/products/solar-cycle-progression}}
during the month of June 2025 was 131\,SFU, which should be contrasted
with 70\,SFU during solar minimum. Next, specifically on 20 June,
the Sun emitted an X1.9 solar flare early in the (UT) morning of
20
June\footnote{\url{https://science.nasa.gov/blogs/solar-cycle-25/2025/06/20/strong-flare-erupts-from-sun-5/}}.

From Figure~\ref{fig:HeI_2025_06_20} we see that there is significant
variability during the passage across the SEP region.  Some of this
is due to SPHEREx pointing at different locations (all in the
vicinity of the terminator plane). In fact, the models of \cite{stw+15}
show significant spatial structures in the polar regions.  A novel
fish-eye lens imager with a narrow band centered on 1.0833\,$\mu$m,
recently commissioned at the Poker Flat Research Range (65$^\circ$N,
147$^\circ$W; \citealt{tmu+24}), show large-scale structures in the
images \citep{mcg+24}.

Some of the variations could be due to an increase in metastable
helium resulting from intermittent electron precipitation. This
phenomenon has been seen in a 17-night winter campaign undertaken
at the Kjell Henriksen Observatory, Svalbard ($78^\circ$N,
$16^\circ$E) by \cite{nkb+25}.  These authors performed spectral
imaging observations covering the wavelength range 1.06--1.13\,$\mu$m
along with an incoherent scatter radar facility (EISCAT Svalbard
Radar).  The near-IR spectra are contaminated by strong and numerous
OH lines.  The authors modeled the spectra and infer emission from
He~I\,1.0833\,$\mu$m.  They find correlation between enhanced
1.0833\,$\mu$m and precipitation of electrons into the thermosphere.

\section{Way Forward}
	\label{sec:WayForward}

The detection of strong He~I\,1.0833\,$\mu$m emission from the
thermosphere, while admittedly a hindrance for astronomers, nevertheless
provides a new diagnostic tool for atmospheric scientists.  The
{\it data from just one orbit} of SPHEREx (see
Figure~\ref{fig:HeI_2025_06_20}) were sufficient to
demonstrate the winter helium bulge phenomenon: a notably higher
abundance of helium in the thermosphere above the wintry pole
relative to that at the summer pole.  Helium is nonreactive, so the
explanation for the bulge is not connected with any photochemical
reactions.  The primary reason is large-scale circulation and
atmospheric dynamics.  The principal physical processes are advection
(specifically, vertical winds) and molecular diffusion (recall that
helium is lighter than oxygen and nitrogen).  The source of energy
for large-scale circulation is the heating of one hemisphere (during
the summer appropriate to that hemisphere) relative to the other
hemisphere; see \cite{lwt+14} for a recent review.

Bearing this summary in mind, we now highlight the possible
contribution of SPHEREx to aeronomy.  SPHEREx data, although confined
to a band of sky close to the terminator plane, are global: a
$3^\circ\times 360^\circ$ band along the terminator plane is covered
in a single 97-minute orbit.  Rigorous testing of global circulation
models and their assumptions of key physical processes can be
undertaken by pitting the data provided by SPHEREx against the model
predictions. Were SPHEREx to be funded to operate beyond the nominal
2-year prime phase, then the impact of solar activity on the helium
bulge could be quantified.  

Finally,  as stressed by \cite{stw+15}, the vertical distribution
of helium has a direct effect on the orbital longevity of low-earth
orbit (LEO) satellites.  Given the increasing number of LEO satellites,
it is of practical benefit to continue studying helium in the
thermosphere.  Most of the satellites are located below 650\,km
(e.g., Starlink is located at about 550\,km, as is the case for
Hubble Space Telescope and Swift Observatory).  Earth observation
satellites and defense satellites tend to above 600\,km.

Bearing this in mind, the reader should be aware that the column
density of metastable helium atoms measured by SPHEREx is the column
starting at 650\,km and going to higher heights. In contrast,
ground-based observations of He~I emission yield the complete column
density of metastable helium.  The nightly monitoring during the
winter season of Svalbard (for example \citealt{nkb+25}) can be
compared against SPHEREx data collected over about 15 passes per
day.  Coordinated lidar observations (e.g., \citealt{gk25}) would
be particularly exciting, since these observations directly yield
densities at the height of SPHEREx.  In addition, adding electron
density profile measurements with incoherent scatter radar observations
may inform us of the origin of rapid brightening of the helium line.

One of the goals of SPHEREx is to use the technique of intensity
mapping to investigate the early universe. This technique is a
statistical technique that rests on the assumption that the
non-astronomical signals are thoroughly understood. SPHEREx, being
in a low-earth orbit, will suffer from the terrestrial foreground
of which the most prominent feature is the He~I\,1.0833\,$\mu$m
line.  As noted in the SPHEREx {\it Supplement} (cf.\ see also
Figures~\ref{fig:NRLMSIS} and \ref{fig:IRI_Hep}) there will be
fainter lines from atomic neutral and ionized species.  These line
features will vary, for a variety of reasons, including the activity
of the sun.  So, astronomers interested in the edge of the Universe
may find themselves unwittingly in a position of precision
modeling the very
near foreground.

In \S\ref{sec:SolarActivity} I noted that we are now past the solar
maximum.  So, in the future, as the sun quietens down, the emissions
from the atmosphere can also be expected to decrease.  It is
anticipated that solar minimum will be reached in 2030. There is
considerable value in operating SPHEREx through solar minimum.

\acknowledgements

I thank Kishalay De (Columbia University) for a short tutorial on
SPHEREx data and for several discussions. I am grateful to E. Sterl
Phinney (Caltech) for patiently clarifying the definition of satellite
inclination, Mike Werner (Jet Propulsion Laboratory, Pasadena) and
Stefan Noll (German Aerospace Center, Bavaria, Germany) for detailed
feedback on the manuscript, and Guilio Del Zanna (Cambridge University,
Cambridge, UK) for supplying the model bound-free opacity table
for metastable helium.  I thank Sean Bryan (Arizona State University,
Tempe, Arizona) and Jamie Bock (Caltech) for discussions regarding
SPHEREx, Bruce Draine (Princeton University) for alerting me about
the \cite{A49} paper, and Lynne Hillenbrand (Caltech) and Tom Greene
(IPAC) for feedback.

I am grateful to John Meriwether (Clemson University, South Carolina;
Center for Solar-Terrestrial Research, New Jersey Institute of
Technology), Takanori Nishiyama (National Institute of Polar Research,
Tachikawa, Tokyo Prefecture, Japan), Christopher Geach (Deutsches
Zentrum f\"ur Luft- und Raumfahrt, Oberpfaffenhofen, Bavaria,
Germany) and Edwin Mierkiewicz (Embry-Riddle Aeronautical University,
Daytona Beach, Florida) for patiently answering my questions about
basic aeronomy terminology and clarifying various questions about
helium in the thermosphere.

This publication makes use of data products from the Spectro-Photometer
for the History of the Universe, Epoch of Reionization and Ices
Explorer (SPHEREx), which is a joint project of the Jet Propulsion
Laboratory and the California Institute of Technology, and is funded
by the National Aeronautics and Space Administration.\\

\bibliography{bibmHeliumSPHEREx}{}
\bibliographystyle{aasjournal}

\appendix

\section{Data Analysis}
	\label{sec:DataAnalysis}

\cite{cwa+20} is a good starting point for a reader to learn about
the mission.  As noted in \S\ref{sec:Observations} the fundamental
reference to SPHEREx data is the ``Explanatory Supplement", provided
by IRSA.

SPHEREx has a fixed observation pattern. The telescope is pointed
in a fixed direction (in the general vicinity of the terminator
plane) for a duration of about 113.6\,s. The six detectors are
readout, the telescope slews to the next programmed position,
and the sequence is repeated.  Here we focus on band~1 data.

\begin{figure}[htbp]  
 \centering
   \includegraphics[width=2.7in]{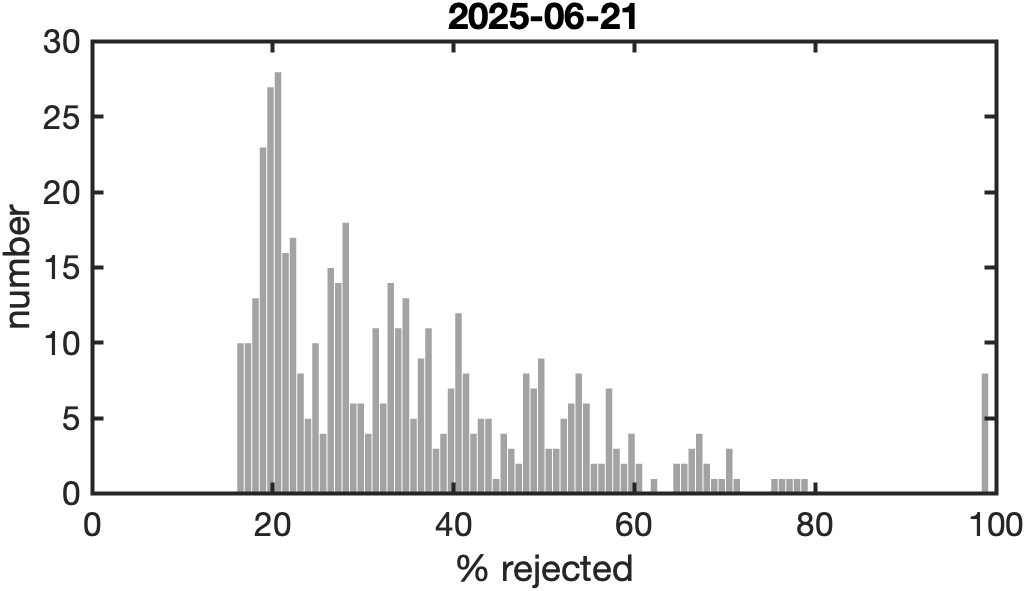}\qquad\qquad
    \includegraphics[width=2.7in]{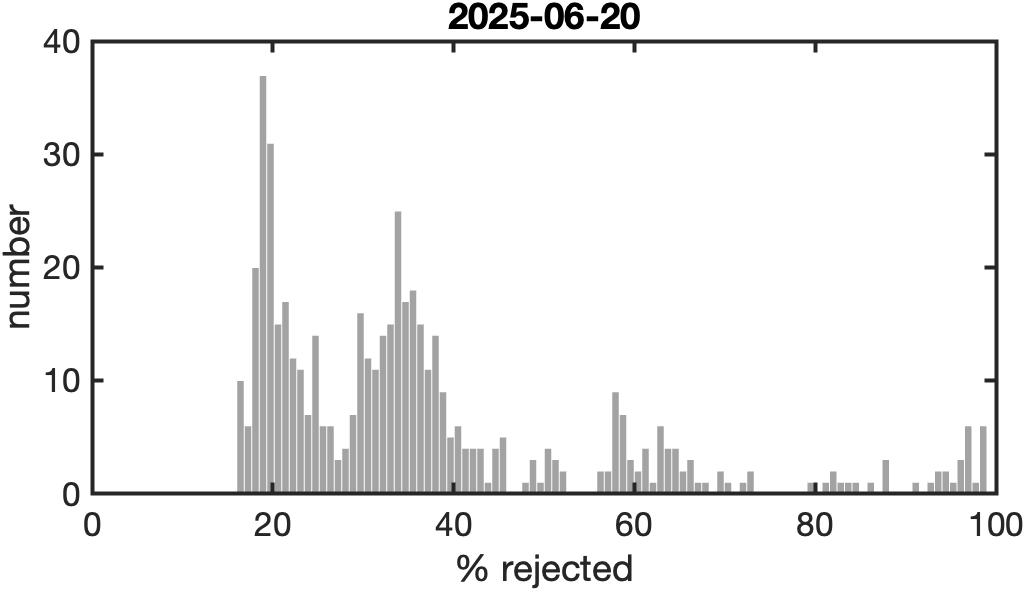}
     \caption{\small Histogram of percentage of rejected pixels.
     For each band 1 frame, the number of pixels which are flagged
     are counted. Dividing this by the total number of pixels yields
     the fraction of pixels flagged ($x$-axis).  The $y$-axis is the
     number of band 1 image frames.  The UT day for each data set
     is shown in the title. }
 \label{fig:histogram_frej}
\end{figure}

The data product is a FITS file with seven extensions. Of relevance
to us are the following: (1) a $2040\times 2040$-pixel ``spectral
image" (see Figure~\ref{fig:band1_image})), (2) a $2040\times
2040$-pixel flag matrix, (3) a $2040\times 2040$-pixel variance
matrix and (4) a $2040\times 2040$-pixel image of the zodiacal light
as estimated from a preflight model.  Some of these flags are
generated at Level 1 (e.g., known non-functional/bad pixels) and
others generated at Level 2 (e.g., charge spillover, cosmic rays,
saturation, missing data, hot pixel, cold pixel, persistence,
outlier). Passages through the South Atlantic Anomaly (SAA) are
expected to suffer severely from cosmic-ray impacts.  The reader
should note that the flag with the key word ``SOURCE" is applied
for pixels assigned to a known source.  So, fields at low Galactic
latitudes will suffer from a high rate of rejection. In our analysis,
given our focus on obtaining the sky spectrum, we rejected all
pixels which were flagged. The histogram of the fraction of rejected
pixels is summarized in Figure~\ref{fig:histogram_frej}.

Separately,  IRSA supplies a matrix whose elements are the nominal
wavelength associated with the corresponding pixel in the spectral
image. I assigned these nominal wavelengths to each of the unflagged
pixels of the spectral image. The pixels in the spectral image are
then sorted by wavelength.  The result is a 1-D spectrum.  This
spectrum is broken up into ``sub-bands" of 701 pixels (``segment").
The median of each segment is obtained. The formal wavelength of
each segment is the arithmetic mean of the wavelengths in each
segment.  The sky spectrum is the run of this median with wavelength.
The He~1.0833\,$\mu$m line integrated intensity was obtained by
summing the spectrum over the wavelength range 1.054--1.106\,$\mu$m
(see Figure~\ref{fig:HeI}).  The typical uncertainty in the integrated
flux of the He~I line is about $3\times 10^{-4}\,{\rm MJy\,ster^{-1}\,\mu
m}$. The zodiacal continuum is obtained as the mean value in the
wavelength range $[1.0,1.2]\,\mu$m.

The He~I line has a small dip close to the peak (Figure~\ref{fig:HeI}).
This feature is seen in almost all datasets.  The feature seems to
be an anticipated artifact of the system (see Figure~18 of
\citealt{cbb+25}). The broad hump at 0.85\,$\mu$m has been identified
by the PI team as a leakage of the He~I\,1.0833\,$\mu$m line
(p.\ 16 of the {\it Supplement}).

\section{Lomb-Scargle Analysis}
	\label{sec:Lomb-Scargle}

\begin{figure}[htbp] 
 \centering
  \includegraphics[width=2.7inch]{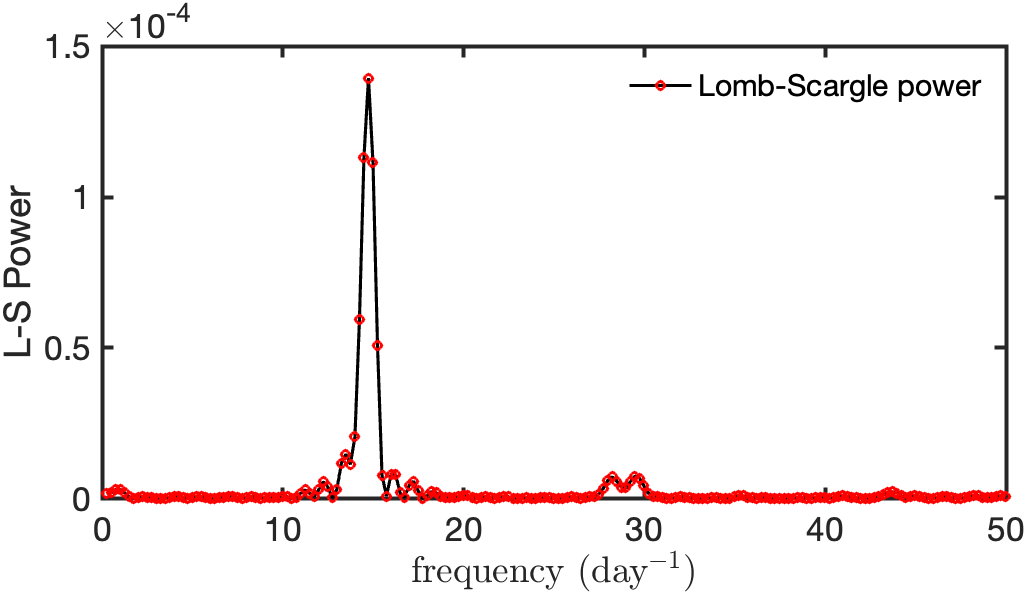}\qquad
  \includegraphics[width=2.7inch]{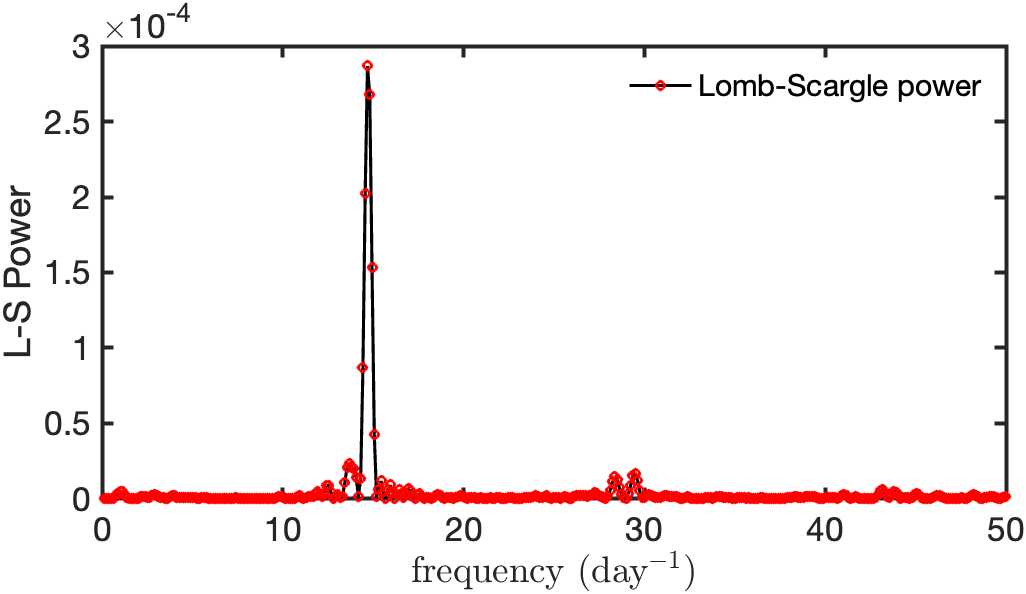}
   \caption{\small Lomb-Scargle periodogram of He~I intensity (left:
   June 20, 2025; right: June 20--21, 2025).  Notice the strong
   fundamental and even structure around the second harmonic.  The
   inverse period from the left panel is $14.739\pm 0.499\,{\rm
   day^{-1}}$ (half-width at half maximum) and that from right panel
   is $14.683\pm 0.251\,{\rm day^{-1}}$.  This corresponds to a
   period of $98.1\pm 1.7\,$minutes. } 
 \label{fig:LombScargle}
\end{figure}

The Lomb-Scargle periodogram of the light curves shown in
Figure~\ref{fig:HeI_2025_06_20} is shown in Figure~\ref{fig:LombScargle}.
From this I infer an orbital period of $98.1\pm 1.7\,$minutes.
According to {\it ISS
Tracker}\footnote{\url{https://isstracker.pl/en?satId=63182}} the
current orbital period of SPHEREx is 97.69\,minutes.

\section{Physical Processes}
	\label{sec:PhysicalProcesses}

\subsection{Photoionization}
	\label{sec:Photoionization}

A rough estimate of the bound-free photo-ionization cross-section
can be obtained by assuming $\sigma_{\rm pi}(\nu)\propto \nu^{-3}$
for $\nu>\nu_{\rm pi}$ along with the normalization, $\int \sigma_{\rm
pi}(\nu)d\nu=\pi e^2/(m_ec)f_{\rm pi}$ \citep{W34}.  $f_{\rm pi}$
is computed using the Thomas-Reiche-Kuhn sum rule.  From
Table~\ref{tab:mHe_lines} we see that the sum of the oscillator
strengths for all upward transition from the 1s2s~${\rm ^3S_1}$
state is 0.66.  The oscillator strength of the decay of the ${\rm
^3S_1}$ to the ground state is negligible, $-2.2\times 10^{-14}$.
Application of the sum rule to the 1s2s~${\rm ^3S_1}$ level then
yields  $f_{\rm pi}=0.34$.  This approach has pedagogical value but
is not expected to be accurate (especially at threshold energies).
\cite{N71} reported improved calculations. A more recent calculation
is by N.\ Badnell which was used in  \cite{dsb+20}.\footnote{Del
Zanna was kind enough to provide me the model cross-section data.}

\begin{figure}[htbp]    
 \plottwo{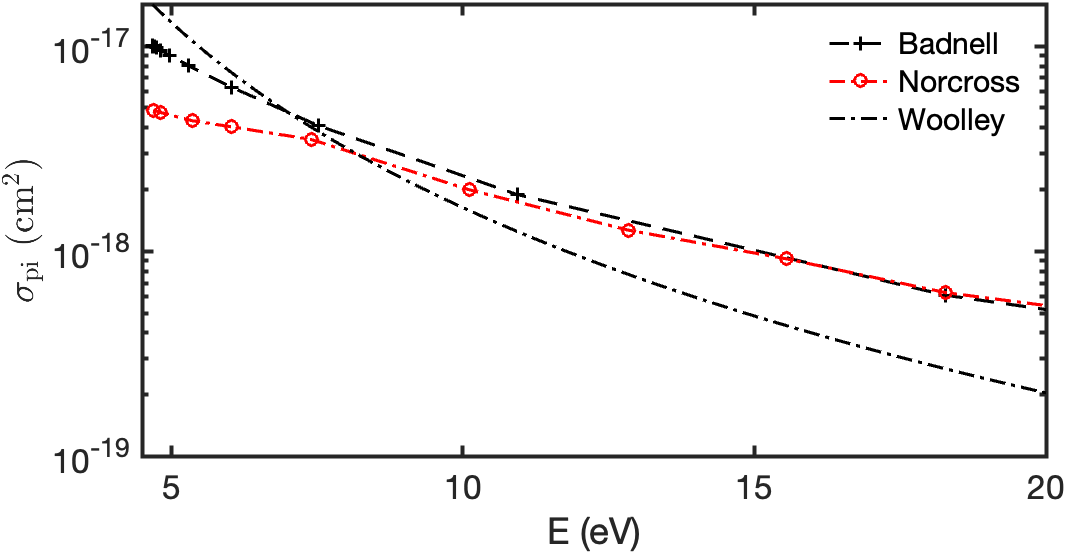}{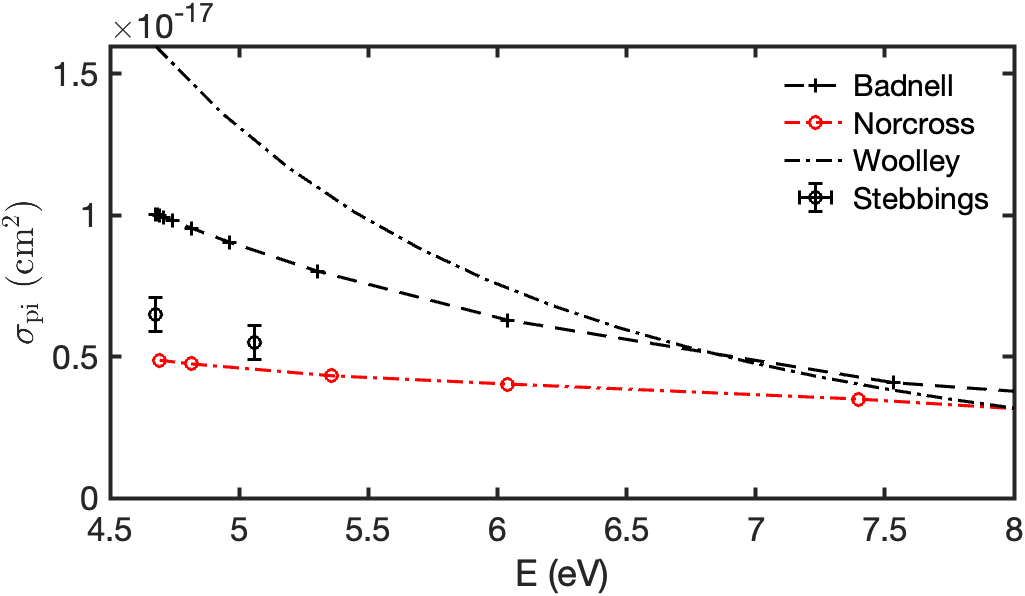}
   \caption{\small (Left): Run of bound-free cross-section  (cm$^2$)
   for metastable helium with photon energy in eV. (Right): Zoom-in
   along with experimental points of  \citet{sdt+73}.  } 
 \label{fig:He_bf}
\end{figure}

The three model calculations are summarized in the left panel of
Figure~\ref{fig:He_bf}. Surprisingly, there is only one experimental
report on this important cross section \citep{sdt+73}. The
measurements are close to threshold and lie between the two model
calculations (right panel of Figure~\ref{fig:He_bf}).

\subsection{Other Processes}
	\label{sec:OtherProcesses}
	
\noindent{\bf Recombination.} The rate of di-electronic recombination
is negligible relative to radiative recombination.  The radiative
recombination coefficients directly to the ground state (1s$^2$)
and to all but the 1s$^2$ level (ground state) are, respectively.
  \begin{eqnarray}
	\alpha_{\rm 1s^2} = 1.54 \times 10^{-13}T_4^{-0.486}\,{\rm
	cm^3\,s^{-1}}, \qquad \alpha_B=2.72 \times
	10^{-13}T_4^{-0.789}\,{\rm cm^3\,s^{-1}}
		\label{eq:Recombination}
 \end{eqnarray}
(\S14.3  of \citealt{D11}).  For the former, the free-bound photon
emitted upon recombination, if absorbed locally, will result in
another ionization. We assume that all ionizing photons escape from
the thermosphere, in which case the recombination coefficient is
$\alpha$, the sum of the two coefficients. The recombination rate
per He$^+$ ion is $\alpha n_e$ where $n_e$ is the electron density.\\

\noindent{\bf Collisional Ionization (Penning Ionization).} A helium
atom in the metastable state is 19.18\,eV  above the ground state.
An encounter with a hydrogen atom can result in the ionization of the
hydrogen and the liberation of an electron.  Another channel, curiously,
results in the formation of {\it helonium} (HeH$^+$;  the very first
molecule formed in the Universe):
 \begin{eqnarray}
	{\rm He (1s2s\ ^3S_1)+H} \rightarrow {\rm HeH^+} + e^-,
	&\qquad&
		{\rm He (1s2s\ ^3S_1)+H} \rightarrow {\rm He} +
		{\rm H^+} + e^- \ .
 \end{eqnarray}
Over a temperature range of 300\,K to 4,500\,K, the sum
of the collisional ionization coefficient for the two reactions is
$Q=5\times 10^{-10}\,{\rm cm^3\,s^{-1}}$.\\

\noindent
{\bf Collisions with electrons.} In astronomical atomic physics it
is traditional to apply the formalism of electron-ion collisions
to all collisions (see Chapter 2 of \citealt{D11}).  The de-excitation
rate coefficient, $\langle\sigma v\rangle_{ul}$, and its inverse
are given by
 \begin{eqnarray}
   \langle \sigma v\rangle_{ul} = \frac{h^2}{(2\pi
   m_e)^{3/2}}\frac{1}{(kT)^{1/2}} \frac{\Omega_{ul}(T)}{g_u}=
   \frac{8.629\times 10^{-8}}{\sqrt{T_4}}\frac{\Omega_{ul}}{g_u}\,{\rm
   cm^{-3}\,s^{-1}}, &\qquad&
     \langle\sigma v\rangle_{lu} =\frac{g_u}{g_l}\sigma_{ul}e^{-E_{ul}/kT}
     \nonumber
 \end{eqnarray}
where $u$ and $l$ stand for upper and lower, $g_u$ is the degeneracy
of the upper level, $\Omega_{ul}$ is the dimensionless collisional
strength, and $E_{ul}$ is the energy difference between the lower
and upper states, respectively.  For electron-ion collisions
$\Omega_{ul}$ has virtually no dependence on $T$.

Collisions that excite metastable helium to any level within the
triplet family lead to a radiative cascade that will eventually
bring the atom back to 1s2s~${\rm ^3S_1}$. In other words, such
collisions do not change the density of metastable atoms.  In the
absence of a strong external source of light (e.g., the sun)
excitations within the triplet family can be important (e.g., in
planetary nebulae). In contrast, collisions which convert a metastable
helium atom to para-helium will result in rapid radiative decay to
the ground state.  Given the modest electron temperatures, we only
consider low energy ($\sim 1\,$eV) excitations which are summarized
in Table~\ref{tab:He_collisions}.

\cite{bbf+00} present computed collisional strengths.  The authors
warn the reader ``Because of the uncertainty that pseudo-resonances
introduce into our cross sections at energies close to threshold,
we limit the low temperature end of our tabulation to about 6,000
degrees." We use the entries for 5,600\,K.  The collisions under
discussion are electron \&\ neutral.  Thus, the collisional strength
at, say, 3,000\,K, will be lower. We reduce the strengths by 20\%
from that given in Table~\ref{tab:He_collisions}.

 \begin{deluxetable}{lrr}[htb]
\label{tab:He_collisions}
 \tablecaption{Collisional de-excitations of He by electrons}
 \tablewidth{0pt} \tablehead{ 
 \colhead{state} &
 \colhead{$E_{ul}$ (eV)} & 
 \colhead{$\Omega_{ul}$} }
\startdata 
1s2p ${\rm ^3P_{0,1,2}}$  & 1.1438     &15.1 \\
1s2s ${\rm ^1S_0}$            & 0.7962    & 2.4\\ 
1s$^2$\ ${\rm ^1S_0}$         & $19.8196$ &0.06\\ 
\enddata 
 \tablecomments{ For the first two entries the lower state is
 1s2s~${\rm ^3S_1}$ while that for the third 1s2s~${\rm ^3S_1}$
 is the upper state.  The value of $\Omega_{ul}$ in column 3 is
 that at $T=5,600\,$K (from \citealt{bbf+00}).}
\end{deluxetable}

\end{document}